\documentclass[twocolumn,showpacs,preprintnumbers,amsmath,amssymb,prb]{revtex4}

\usepackage{graphicx}
\usepackage{color}

\begin{document}

\title{Path-integral simulation of graphene monolayers under tensile stress}
\author{Carlos P. Herrero}
\author{Rafael Ram\'irez}
\affiliation{Instituto de Ciencia de Materiales de Madrid,
         Consejo Superior de Investigaciones Cient\'ificas (CSIC),
         Campus de Cantoblanco, 28049 Madrid, Spain }
\date{\today}

\begin{abstract}
Finite-temperature properties of graphene monolayers under tensile 
stress have been studied by path-integral molecular dynamics (PIMD) 
simulations. This method allows one to consider the quantization of 
vibrational modes in these crystalline membranes and to analyze 
the influence of anharmonic effects in the membrane properties.
Quantum nuclear effects turn out to be appreciable in structural
and thermodynamic properties of graphene at low temperature, and
they can even be noticeable at room temperature.
Such quantum effects become more relevant as the applied stress
is increased, mainly for properties related to out-of-plane atomic 
vibrations.
The relevance of quantum dynamics in the out-of-plane motion 
depends on the system size, and is enhanced by tensile stress.
For applied tensile stresses, we analyze the contribution of the
elastic energy to the internal energy of graphene.
Results of PIMD simulations are compared with calculations based
on a harmonic approximation for the vibrational modes of the 
graphene lattice. This approximation describes rather well 
the structural properties of graphene, provided that the frequencies of 
ZA (flexural) acoustic modes in the transverse direction include a
pressure-dependent correction.
\end{abstract}



\maketitle

\section{Introduction}

In recent years there has been a surge of interest on two-dimensional
materials, and graphene in particular, due to their unusual electronic, 
elastic and thermal properties.\cite{ge07,ca09b,fl11,ro17}
In fact, graphene displays high values of thermal
conductivity,\cite{gh08b,ni09,ba11} as well as large in-plane elastic
constants.\cite{le08} Its mechanical properties are also important for 
possible applications, such as cooling of electronic devices.\cite{se10,pr10}

The structural arrangement for pure defect-free graphene
corresponds to a planar honeycomb lattice. At finite temperatures, 
there appear out-of-plane displacements of the C atoms, 
and for $T \to 0$, quantum fluctuations related to zero-point
motion give rise to a departure of strict planarity of the graphene 
sheet.\cite{he16}
In particular, one has low-lying vibrational excitations associated to
large-scale ripples perpendicular to the plane.\cite{fa07}
Moreover, a graphene sheet can actually bend and depart from planarity 
for other reasons, such as the presence of defects and external 
stresses.\cite{me07,an12}

  A thin membrane crumples in the presence of a compressive stress. 
This behavior has been investigated during the last three decades 
in lipid membranes\cite{ev90,sa94} and polymer films.\cite{ce03,wo06}
In graphene, crumpling originates from out-of-plane phonons as
well as from static wrinkling, and has been observed in both
supported and freestanding samples.\cite{ki11,ni15}
Mechanical properties such as stiffness and bending rigidity 
can be renormalized due to crumpling.\cite{ru11,ko13,ko14}
For graphene, it has been found that the maximum compressive stress that 
a freestanding sheet can sustain without crumpling decreases with system 
size, and has been estimated to be about 0.1 N/m at room temperature
in the thermodynamic limit.\cite{ra17}

A tensile stress applied in the graphene plane does not affect the
planarity of the sheet, but causes appreciable changes in the elastic 
properties of the material.\cite{lo17} Thus, it has been observed that
the in-plane Young modulus is increased by a factor of three when
applying a stress of 1 N/m.\cite{ra17}
The bending rigidity $\kappa$ does also change with the tensile stress,
and in fact it decreases but not so critically as the in-plane elastic 
constants.
In this context, it is important to note that the actual area per atom, 
$A$, is not readily measurable, and the accessible observable is usually
its projection, $A_p$, onto the mean plane of the membrane ($A_p \leq A$).
Thus, one may refer the elastic properties of graphene either to the
area $A$ or to $A_p$, which may behave in very different ways.
For example, a negative thermal expansion coefficient is found for
graphene when one refers to $A_p$, but the thermal expansion 
associated to the area $A$ is positive.\cite{po11b,he16}

Recent experimental and theoretical work has shown the influence of strain 
in several characteristics of graphene, such as electronic transport, 
optical properties, and the formation of moir\'e patterns.\cite{wo14}
Similar properties have been also studied 
in other two-dimensional materials, as metallic dichalcogenides.\cite{am16}

Equilibrium and dynamical properties of graphene have been studied 
earlier by using Monte Carlo and molecular dynamics simulations. 
These simulations were based on
{\em ab-initio},\cite{sh08,po11b,an12b,ch14}
tight-binding,\cite{he09a,ca09c,ak12,le13}
and empirical interatomic potentials.\cite{fa07,sh13,lo09,ra16,ma14,lo16}
In most of these simulations, C atoms were treated as classical particles,
which is accurate at relatively high temperatures but is not suitable 
to study thermodynamic variables at low temperature.
The quantum character of the atomic motion can be taken into account
by employing path-integral simulations, which allow to consider
quantum and thermal fluctuations in many-body systems
at finite temperatures.\cite{gi88,ce95}
Path-integral simulations of a single graphene sheet have been
lately performed to study equilibrium properties
of this material.\cite{br15,he16}
Moreover, nuclear quantum effects have been studied 
by a combination of density-functional theory and a quasi-harmonic
approximation for vibrational modes in this crystalline
membrane.\cite{mo05,sh12}

In this paper, we employ path-integral molecular dynamics (PIMD) 
simulations to study structural and vibrational properties of graphene
under tensile stress.
We consider different sizes for the simulation cell, as finite-size 
effects are known to be important for some properties of 
graphene.\cite{ga14,lo16,he16}
The magnitude of nuclear quantum effects in the graphene properties 
is assessed by comparing the results of PIMD simulations with data 
obtained from classical simulations. 
We find that quantum effects are relevant to describe the temperature
and pressure dependence of graphene's real and in-plane areas, as
well as to describe the amplitude of the out-of-plane motion, especially
at low temperatures. 
Our data indicate that the relevance of nuclear quantum effects 
increases as tensile stress is raised.
Results of PIMD simulations are compared with data derived from a 
harmonic approximation for the out-of-plane vibrations. This approximation
turns out to be rather accurate, provided that the vibrational
frequencies of ZA acoustic modes are conveniently renormalized 
for different applied stresses. 

The paper is organized as follows. In Sec.\,II, we present the
computational method used in the simulations.  Structural properties such 
as in-plane $A_p$ and real area $A$ are given in Sec.~III as a function 
of applied stress.
Results for the internal, vibrational, and elastic energy of graphene 
are discussed in Sec.~IV.
In Sec.~V we study the out-of-plane atomic motion, with emphasis on the 
competition between classical-like and quantum dynamics. 
In Sec.\,VI we summarize the main results.

\section{Computational Method}

We use the PIMD method to obtain equilibrium properties of graphene
under tensile stress.
This procedure is based on the Feynman path-integral formulation of 
statistical mechanics, a nonperturbative technique to study many-body
quantum systems at finite temperatures.\cite{fe72}
The implementation of this formulation for numerical simulations is based 
on an isomorphism between the quantum system and a fictitious classical 
system, in which each quantum particle is described by a ring polymer 
(corresponding to a cyclic quantum path) composed of  $N_{\rm Tr}$ 
(Trotter number) {\em beads}.\cite{kl90} This becomes exact in the limit 
$N_{\rm Tr} \to \infty$. 
Details on this simulation technique can be found
elsewhere.\cite{ch81,gi88,ce95,he14}
The dynamics in PIMD is artificial, since it does not correspond 
to the actual dynamics of the real quantum particles.
However, it is useful for sampling the
many-body configuration space, yielding accurate results for 
time-independent equilibrium properties of the actual quantum system.

The Born-Oppenheimer surface for the nuclear dynamics is derived here
from an effective empirical potential, developed for carbon-based systems, 
namely the so-called LCBOPII.\cite{lo05}
This is a long-range carbon bond order potential, which was previously 
used to perform classical simulations of diamond,\cite{lo05} 
graphite,\cite{lo05} liquid carbon,\cite{gh05} 
as well as graphene sheets.\cite{za09,fa07,lo16}
A relevant application of this effective potential was the calculation
of the carbon phase diagram including diamond, graphite, and the liquid, 
and showing its precision by comparison of the
predicted diamond-graphite line with experimental results.\cite{gh05b}

The LCBOPII potential has been more recently employed to study graphene,
giving a good description of elastic properties such as the Young's
modulus.\cite{za09,po12} 
According to previous simulations,\cite{ra16,he16,ra17} the original LCBOPII
parameterization has been slightly modified to increase the zero-temperature
bending constant of graphene from 1.1 eV to a value of 1.49 eV, more
consistent with experimental data.\cite{la14}
This effective potential was lately used to perform PIMD simulations, 
allowing to assess the extent of quantum effects in graphene sheets 
from a comparison with results of classical simulations.\cite{he16}

Other effective interatomic potentials have been employed in recent
years to study various properties of graphene. In particular, the
AIREBO potential model has been widely used 
in this field.\cite{sf15,me15,zo16,an16,gh17}
Comparing the LCBOPII and AIREBO models, we find that they yield
very similar equilibrium C--C distance and in-plane thermal expansion
coefficient, as derived from classical molecular dynamics 
simulations.\cite{me15,gh17,he16}
Results for the Young's modulus of graphene derived from the
LCBOPII potential are closer to those given by {\em ab initio}
calculations.\cite{me15}

The calculations presented here were carried out in 
the isothermal-isobaric ensemble,
where we fix the number of carbon atoms ($N$), the applied stress ($P$), 
and the temperature ($T$).
We employed effective algorithms for carrying out PIMD simulations
in this statistical ensemble, as those presented in the
literature.\cite{tu92,tu98,ma99,tu02}  Specifically,
we used staging variables to define the bead coordinates, and
the constant-temperature ensemble was achieved by coupling chains
of four Nos\'e-Hoover thermostats. 
A supplementary chain of four barostats was coupled to the area 
of the simulation box to give the required pressure $P$.\cite{tu98,he14}
The kinetic energy $K$ has been calculated by using the so-called virial
estimator, which has a statistical uncertainty smaller than the 
potential energy, $V$, of the system.\cite{he82,tu98}
Other technical details about the simulations presented here are the
same as those given elsewhere.\cite{he06,he11,ra12}

Atomic forces were analytically derived from position derivatives of the
instantaneous potential energy $U$ (note that $V = \langle U \rangle$). 
The estimator of the two-dimensional (2D) stress tensor ${\bf \tau}$ is 
the same as that employed in previous works,\citep{ra16,ra17}
and its formulation for PIMD simulations of graphene is similar to that
given earlier for three-dimensional (3D) solids.\cite{ra08,he14}
Details on the pressure estimator employed here are presented in 
Appendix~\ref{app1}.
The mechanical stress $P$ in the $(x, y)$ plane of graphene is obtained
from the trace of the tensor ${\bf \tau}$:
\begin{equation}
  P = \frac{1}{2} \left( \tau_{xx} + \tau_{yy} \right)  \, .
\end{equation}
Note that in the case of applying a large compressive stress 
($P > 0$, not considered here), one may have severe bending or crumpling 
of the graphene sheet. In this case the in-plane stress $P$ may 
appreciably differ from the actual stress felt by the {\em real}
graphene surface (related to the area $A$).\cite{ra17,fo08}

We consider rectangular simulation cells with similar side lengths
$L_x$ and $L_y$ in the $x$ and $y$ directions of the reference
plane, and periodic boundary conditions were assumed.
Sampling of the configuration space has been carried out at temperatures
between 12~K and 2000~K.
For comparison with results of PIMD simulations, some classical
molecular dynamics (MD) simulations have been also performed.
In our context this is achieved by by setting $N_{\rm Tr}$ = 1.
For the quantum simulations, $N_{\rm Tr}$ was taken 
proportional to the inverse temperature: $N_{\rm Tr} \, T$ = 6000~K, 
which roughly gives a constant precision in the PIMD results at different 
temperatures.\cite{he06,he11,ra12}
Cells of size up to 8400 and 14720 atoms were considered for PIMD 
and classical MD simulations, respectively.
For a given temperature, a typical simulation run consisted of
$3 \times 10^5$ PIMD steps for system equilibration, followed by
$4 \times 10^6$ steps for the calculation of ensemble average properties.

\section{Structural properties}

The simulations presented here were performed in the
isothermal-isobaric ensemble, as explained above in Sec.~II.
Thus, in a simulation run we fix the number of carbon atoms $N$,
the temperature $T$, and the applied stress $P$ in the $(x, y)$ plane,
allowing for changes in the in-plane area of the simulation
cell for which periodic boundary conditions are assumed.
Carbon atoms are free to move in the out-of-plane direction
($z$ coordinate), and in general any measure of the {\em real} surface
of a graphene sheet at $T > 0$ should give a value larger than
the area of the simulation cell in the $(x, y)$ plane.
In this line, there has appeared in recent years a discussion in the
context of biological membranes, dealing with the question whether
it is more convenient to describe the properties of those membranes 
using the concept of a real surface rather than a {\em projected} (in-plane)
surface.\cite{im06,wa09,ch15}  
The same question has been also recently raised for crystalline membranes 
such as graphene.\cite{po11b,he16,ni17,ra17}
This can be important for addressing the calculation of thermodynamic 
properties, because the in-plane area $A_p$ is the variable conjugate 
to the stress $P$ used in our simulations, and the real area $A$
(also called effective, true, or actual area in the 
literature\cite{im06,fo08,wa09,ch15}) is conjugate to the usually-called 
surface tension.\cite{sa94}
The difference $A - A_p$ has been recently denoted as {\em hidden} area by 
Nicholl {\em et al.}\cite{ni17}

\begin{figure}
\vspace{-1.0cm}
\includegraphics[width=8.5cm]{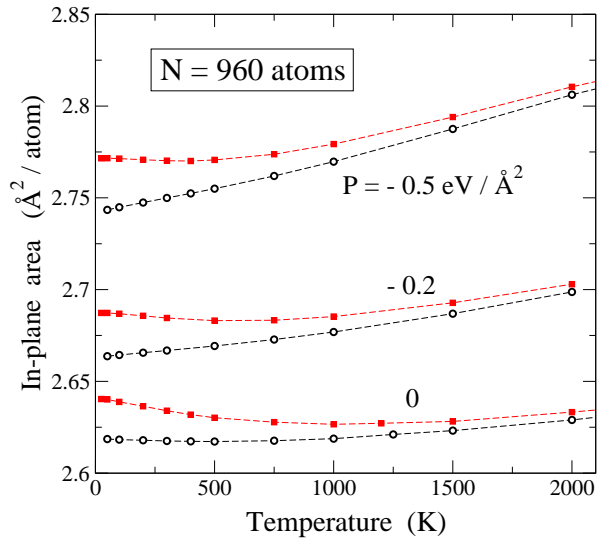}
\vspace{-0.5cm}
\caption{In-plane area $A_p$ vs. temperature for graphene, as derived
from classical (open circles) and PIMD simulations (solid squares)
for $N$ = 960 and different stresses.
From bottom to top: $P$ = 0, $-0.2$, and $-0.5$ eV \AA$^{-2}$.
Error bars are less than the symbol size.
Lines are guides to the eye.
}
\label{f1}
\end{figure}

The real area $A$ in 3D space is calculated here by a triangulation 
based on the atomic positions along a simulation run.
$A$ is obtained from the areas associated to $N/2$ structural hexagons.
Each hexagon contributes by a sum of six triangles, each one formed 
by the positions of two neighboring carbon atoms and the barycenter 
of the hexagon (mean point of its six vertices).\cite{ra17}
There are other similar definitions that can be employed for the area $A$, 
as those based on the interatomic distance C--C.\cite{ha16,he16}
The area $A$ based on triangulation employed here seems more precise 
to deal with the 2D nature of a graphene layer in 3D space.
It has been shown earlier that $A$ has a very small size effect,
in fact negligible in comparison with that appearing 
for $A_p$.\cite{ra17}

In Fig.~1 we show the temperature dependence of the in-plane area 
$A_p$, obtained from classical MD (open circles) and PIMD simulations 
(solid squares) for a supercell with $N$ = 960 atoms.
Results are given for $P$ = 0, $-0.2$, and $-0.5$ eV \AA$^{-2}$.
Tensile stress causes not only an increase in $A_p$, but
its temperature dependence also changes.
For each considered value of the stress, the curve $A_p(T)$ 
derived from quantum simulations displays a minimum, 
that shifts to lower temperatures as the tensile stress is increased. 
Thus, such a minimum evolves from $T_m \approx$ 1000~K
for $P = 0$ to $\approx$ 400~K for $P = -0.5$ eV \AA$^{-2}$. 
In the classical simulations, however, one finds a shallow minimum for
$P = 0$, that is absent for the tensile stresses shown in Fig.~1
(in fact we did not observe it for $P = -0.1$ eV \AA$^{-2}$ either,
not shown in the figure).
The classical results for $P = 0$ are similar to those found in earlier 
classical Monte Carlo and MD simulations of graphene single 
layers.\cite{za09,ga14,br15}

At low $T$ the results of PIMD simulations verify $d A_p / d T \to 0$, 
i.e., the corresponding curves shown in Fig.~1 (solid symbols) become flat 
close to $T = 0$, as required by the third law of thermodynamics.
In the limit $T \to 0$, the difference between quantum and classical results
converges to 0.022 \AA$^2$/atom in the absence of applied stress ($P = 0$). 
This difference decreases for rising temperature, as nuclear quantum effects 
become less important. For $P = -0.5$ eV \AA$^{-2}$, we find for 
$T \to 0$ a difference of 0.031 \AA$^2$/atom.
The increase in $A_p$ at low temperature is due to zero-point motion
associated to in-plane acoustic modes (LA and TA). The frequency of
these modes decreases for increasing $A_p$ (i.e., when tensile stress is
increased, according to positive Gr\"uneisen parameters), and therefore 
their vibrational amplitudes are larger. This causes a larger zero-point 
expansion of $A_p$ for larger tensile stress.

\begin{figure}
\vspace{-1.0cm}
\includegraphics[width=8.5cm]{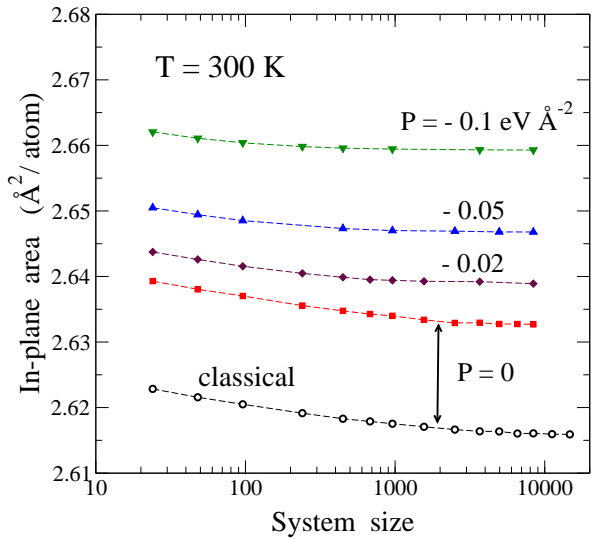}
\vspace{-0.5cm}
\caption{In-plane area $A_p$ vs. system size, as derived from PIMD
simulations for $T$ = 300~K and various tensile stresses (solid symbols).
From top to bottom: $P$ = $-0.1$ (triangles down), $-0.05$ (triangles up),
$-0.02$ (diamonds), and 0 eV \AA$^{-2}$ (squares).
For comparison, open squares indicate results of classical MD simulations
for $P$ = 0.  Error bars are less than the symbol size.
Lines are guides to the eye.
}
\label{f2}
\end{figure}

The presence of a minimum in the $A_p(T)$ curves derived from PIMD
simulations is due to two competing effects, as discussed earlier for 
graphene without stress.\cite{ga14,mi15b,he16}
On one hand, the area $A$ increases as temperature is raised, and
on the other hand, surface bending gives rise to a decrease in its
2D projection, i.e., $A_p$.
At low $T$, this decrease associated to out-of-plane motion dominates
the thermal expansion of the real surface, and $d A_p / d T < 0$.
For the quantum results, the thermal expansion at low $T$ is very small
compared to the classical calculations for which
$\lim_{T \to 0} d A / d T > 0$, thus causing a more appreciable decrease
in $A_p$ for raising $T$ in the quantum case.
At high temperatures, the increase in $A$ predominates over the contraction
in the projected area due to out-of-plane motion.

For unstressed graphene it has been indicated that finite-size effects can 
be important for several structural properties of the crystalline 
membrane.\cite{he16,ra17}
It is now worthwhile to consider finite-size effects for the in-plane 
area of graphene under stress.
In Fig.~2 we present the size dependence of $A_p$ for several 
tensile stresses.
In all cases, $A_p$ decreases for increasing $N$, and reaches a well-defined
plateau for large sizes. One observes that the convergence to the large-size
value is faster for larger tensile stress. Moreover, the difference
between the large-size limit and the value corresponding to $N$ = 24 
(the smallest supercell considered here) appreciably decreases from
$6.7 \times 10^{-3}$ \AA$^2$/atom for $P = 0$ to 
$2.8 \times 10^{-3}$ \AA$^2$/atom for $P = -0.1$ eV \AA$^{-2}$.

For comparison, we also present in Fig.~2 results for $A_p(N)$ derived from
classical MD simulations at 300 K. The difference between quantum and classical 
results for $P$ = 0 amounts to 0.017 \AA$^2$/atom, and it is nearly constant
for the system sizes considered here. 
This difference increases to 0.022 \AA$^2$/atom at $P = 0$ in the
low-temperature limit, as indicated above (see Fig.~1).
For $P = -0.1$ eV \AA$^{-2}$ and $T = 300$~K, our classical simulations
yield an $A_p(N)$ curve similar to the quantum one (not shown in Fig.~2
to avoid overcrowding). In particular, for a system size $N = 960$, we
found an in-plane area $A_p$ = 2.6423 \AA$^2$/atom, so that the difference
between classical and quantum results at this tensile stress is similar 
to that found for $P = 0$.
It is interesting to note that
the increase in area $A_p$ due to quantum nuclear motion at 300 K is 
the same as that caused by a relatively large tensile stress of about 
$-0.07$ eV \AA$^{-2}$ ($\sim -1$~N/m).

\begin{figure}
\vspace{-1.0cm}
\includegraphics[width=8.5cm]{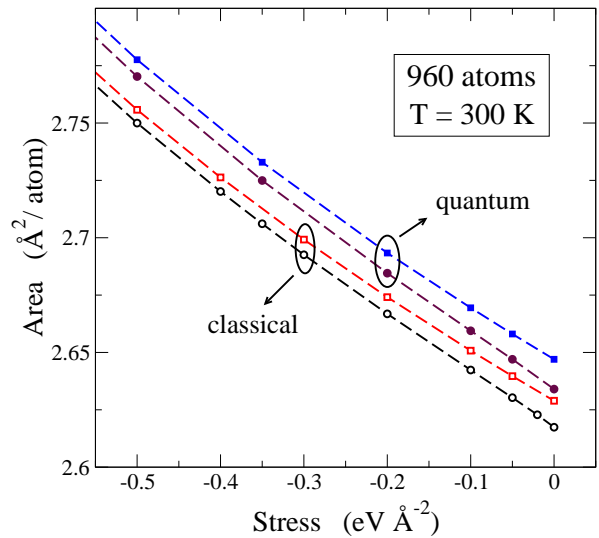}
\vspace{-0.5cm}
\caption{Real area $A$ and in-plane area $A_p$ of graphene vs. tensile stress,
as derived from classical (open symbols) and PIMD simulations (solid symbols)
for $N$ = 960 and $T$ = 300~K.
Squares and circles correspond to $A$ and $A_p$, respectively.
Error bars are less than the symbol size.
Lines are guides to the eye.
}
\label{f3}
\end{figure}

We now turn to the real surface of graphene and its measure through the
area $A$.   It was shown earlier from simulations at $P = 0$ that
the surface $A$ is larger than $A_p$, and the difference between both
increases with temperature. This is clear from the fact that $A_p$ is
a 2D projection of $A$, and the actual surface becomes
increasingly bent as temperature is raised and the amplitude of
out-of-plane atomic vibrations becomes larger.
An important difference between the temperature dependence of $A$ and
$A_p$ is that the latter first decreases for increasing $T$ and then
it increases at higher $T$, with a minimum at a temperature $T_m$.
For rising tensile stress, the vibrational amplitude in the $z$ direction
decreases (see below), so that the temperature $T_m$ of minimum $A_p$ 
is lowered.
This becomes even clearer in the results of classical simulations, for
which the shallow minimum in the curve $A_p(T)$ disappears at relatively
low pressures, and it is not observed in the data presented for $P$ = 
$-0.2$ and $-0.5$ eV \AA$^{-2}$ in Fig.~1.
For the area $A$ one does not observe the decrease displayed
by $A_p$ in both classical and quantum simulations at low temperatures
(see Ref.~\onlinecite{he16} for results at $P = 0$).

In Fig.~3 we present the areas $A$ and $A_p$ vs. tensile stress
for a simulation cell including 960 atoms. In both cases, we present
results from classical (open symbols) and PIMD (solid symbols) simulations.
Circles correspond to the in-plane area $A_p$, whereas squares 
represent data for the real area $A$.
One notices that quantum effects are appreciable at room temperature. 
The main aspects of this figure are the following.
Tensile stress causes an increase of about 5\% in both $A$ and $A_p$
from $P = 0$ to $-0.5$ eV \AA$^{-2}$.
Moreover, quantum nuclear effects cause in both cases a surface expansion 
of about 0.02 \AA$^2$/atom, which increases slightly as the tensile stress
is raised.

To make connection of our results derived from atomistic simulations
with an analytical formulation of crystalline membranes,
we note that the relation between $A$ and $A_p$ can be expressed
in the continuum limit (macroscopic view) as\cite{im06,wa09,ra17}
\begin{equation}
  A = \int_{A_p} dx \, dy \, \sqrt{1 + (\nabla h(x,y))^2}  \; ,
\label{aap}
\end{equation}
where $h(x,y)$ is the height of the membrane surface, i.e. the distance to
the reference $(x, y)$ plane.
The difference $A - A_p$ can be calculated in a classical approach by 
Fourier transformation of the r.h.s. of Eq.~(\ref{aap}).\cite{sa94,ch15,ra17}
This requires the introduction of a dispersion relation $\omega({\bf k})$
for out-of-plane modes (ZA band), where ${\bf k} = (k_x, k_y)$ are
2D wavevectors. The frequency dispersion in this acoustic (flexural)
band can be well approximated by the expression
$\rho \, \omega^2 = \sigma k^2 + \kappa k^4$,
consistent with an atomic description of graphene\cite{ra16}
($k = |{\bf k}|$; $\rho$, surface mass density;
$\sigma$, effective stress; $\kappa$, bending modulus).
The effective stress $\sigma$ can be written as $\sigma = \sigma_0 - P$,
with a term $\sigma_0$ that appears at finite temperature even in 
the absence of an applied stress ($P = 0$) due to out-of-plane motion
(at 300 K, $\sigma_0 \approx 6 \times 10^{-3}$ eV \AA$^{-2}$).\cite{ra16}

After Fourier transformation one has for the area per atom:\cite{sa94,ch15}
\begin{equation}
 A = A_p  + \frac {k_B T}{2 N}
      \sum_{\bf k}  \frac {1} {\sigma + \kappa k^2}   \,  .
\label{aap2}
\end{equation}
For large $N$ the sum in Eq.~(\ref{aap2}) can be approximated by 
an integral:\cite{ra17}
\begin{equation}
 A = A_p  \left( 1 + \frac{k_B T}{4 \pi}
      \int_{k_0}^{k_m} d k \frac {k} {\sigma + \kappa k^2} \right)  \,  .
\label{aap3}
\end{equation}
The limits in the integral are the cut-off $k_m = (2 \pi /A_p)^{1/2}$
and the size-dependent minimum wavevector $k_0 = 2 \pi / L$,
with $L = (N A_p)^{1/2}$.
The integral in Eq.~(\ref{aap3}) converges provided that $\sigma > 0$,
which is the case here. It allows us to explicitly write the 
size-dependent ratio $A / A_p$ as
\begin{equation}
 \frac{A}{A_p} = \left( \frac{A}{A_p} \right)_{\infty} - 
          \frac {k_{B} T} {8 \pi \kappa}
          \ln \left(1 + \frac{4 \pi^2 \kappa}{N A_p \sigma} \right) 
\label{aap4}
\end{equation}
with the large-size limit ($N \to \infty$ or $k_0 \to 0$):
\begin{equation}
  \left( \frac{A}{A_p} \right)_{\infty} =
        1 + \frac {k_{B} T} {8 \pi \kappa} 
           \ln \left(1 + \frac{2 \pi \kappa} {\sigma A_p} \right) \, .
\label{aap5}
\end{equation}
Eq.~(\ref{aap4}), although in principle not very accurate for
small system size, yields for $N = 24$, $P = 0$, and $T$ = 300 K 
($\sigma = 6 \times 10^{-3}$ eV \AA$^{-2}$, $\kappa$ = 1.7 eV; 
see Ref.~\onlinecite{ra17}) 
a shift in $A/A_p$ of $-3.1 \times 10^{-3}$, which translates
into an increase in $A_p$ of $8.1 \times 10^{-3}$ \AA$^2$/atom with 
respect to the large-size limit. From the results of our simulations we
find a size effect in $A_p$ of $7.0 \times 10^{-3}$ \AA$^2$/atom 
for $N = 24$.
Note that, apart from the replacement of the sum in Eq.~(\ref{aap2})
by an integral, the above expressions assume harmonic vibrations for 
out-of-plane motion, which becomes less accurate as temperature increases
for the onset of larger anharmonicity.
Note also the appearance of the stress $\sigma$ ($= \sigma_0 - P$) in 
the logarithmic term in Eq.~(\ref{aap4}), which causes that an increase
in tensile stress ($P$ more negative) gives rise to a faster convergence
of the area $A_p$ with system size, according to the results shown
in Fig.~2.

\section{Internal energy}

At $T = 0$ and zero applied stress we find with the LCBOPII potential 
in a classical approach a strictly planar graphene surface with an 
interatomic distance $d_{\rm C-C}$ = 1.4199 \AA, 
i.e., an area of 2.6189 \AA$^2$ per atom, which we call $A_0$.
This corresponds to a graphene sheet with fixed atomic nuclei on their
equilibrium sites without spatial delocalization, giving the minimum
energy $E_0$, taken as a reference for our calculations at
nonzero temperature and applied stress.
In a quantum approach, the limit $T \to 0$ includes
out-of-plane atomic fluctuations associated to zero-point motion,
and the graphene sheet is not strictly planar.
In addition, anharmonicity of in-plane vibrations gives rise to a
zero-point lattice expansion (increase in the area $A$, see Sec.~III), 
which for $T \to 0$ yields an interatomic 
distance $d_{\rm C-C}$ = 1.4287 \AA, around 1\% larger than the classical 
minimum.

\begin{figure}
\vspace{-1.0cm}
\includegraphics[width=8.5cm]{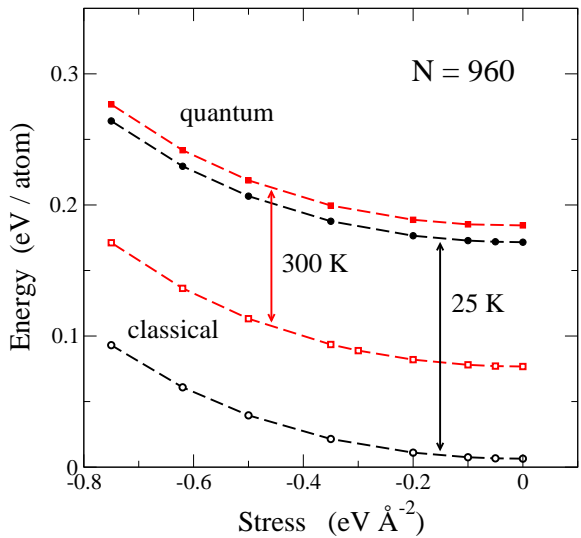}
\vspace{-0.5cm}
\caption{Dependence of the internal energy of graphene on tensile stress,
as derived from classical (open symbols) and PIMD simulations (solid symbols)
at two temperatures: 25 K (circles) and 300 K (squares).
These results were obtained for a simulation cell including 960 carbon atoms.
Vertical arrows indicate the increase in internal energy due to quantization
of nuclear motion at each temperature.
Error bars are less than the symbol size.
}
\label{f4}
\end{figure}

The internal energy $E$ is calculated as a sum of the kinetic $K$ and
potential energy $V$ obtained from the simulations at a given temperature.
In our simulations of graphene, $K$ and $V$ have been
found to slightly increase with system size, and their convergence is
rather fast. Thus, for cells in the order of 200 atoms the size effect 
in the internal energy is almost inappreciable when compared to the largest 
cells.\cite{he16}
The kinetic energy is associated to vibrational motion of carbon atoms
(in-plane and out-of-plane), but the potential energy includes contributions
due to atomic vibrations and to the elastic energy due to changes
in the area $A$ of graphene at finite temperatures and applied stresses
(see below).

In Fig.~4 we display the stress dependence of the internal energy,
$E - E_0$, as derived from classical and PIMD simulations for system size 
$N$ = 960.  Results are shown for $T$ = 25 K (circles) and 300 K (squares). 
Open and solid symbols correspond to classical and PIMD simulations, 
respectively.
At $P = 0$, the classical energy per atom is basically given by the 
vibrational energy $E_{\rm vib}^{\rm cl} = 3 k_B T$, as follows from 
the equipartition theorem in a harmonic approximation (HA).
As the tensile stress is increased ($P$ more negative), the classical
internal energy increases for a given temperature, due to the contribution
of the elastic energy associated to a finite strain in the graphene lattice.
The behavior of the quantum results shown in Fig.~4 is similar to the 
classical ones. The main difference is the increase in internal energy 
caused by quantization of the nuclear motion. For zero stress, this 
increase amounts to 165 meV/atom at 25 K and 107 meV/atom at 300 K.
These shifts do not appreciably change in the stress region considered
here, and in fact the difference between quantum and classical results 
appears to be nearly constant in the results shown in Fig.~4.

To analyze the different contributions to the internal energy, $E(T)$, 
we write
\begin{equation}
    E(T) =  E_0 + E_{\rm el}(A) + E_{\rm vib}(A,T)   \, .
\label{et}
\end{equation}
In this expression $E_{\rm el}(A)$ is the elastic energy corresponding to 
an area $A$, and $E_{\rm vib}(A,T)$ is the vibrational energy of the system.
Although not explicitly indicated, the area $A$ is a function of the
stress $P$ and temperature $T$.
Our simulations give $E(T)$, and using Eq.~(\ref{et}) we can then split 
the internal energy $E(T) - E_0$ into an elastic and a vibrational part.
The vibrational contribution $E_{\rm vib}$ can in turn be split into
kinetic and potential energy parts: $E_{\rm vib} = V_{\rm vib} + K$.

\begin{figure}
\vspace{-1.0cm}
\includegraphics[width=8.5cm]{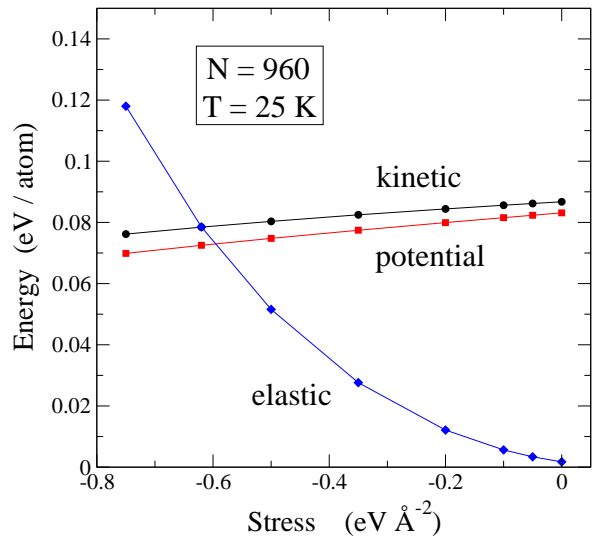}
\vspace{-0.5cm}
\caption{Different contributions to the internal energy of graphene,
as functions of tensile stress. Symbols indicate results of PIMD
simulations for $N$ = 960 atoms and $T$ = 25~K.
Diamonds, elastic energy; circles, kinetic energy;
squares, vibrational potential energy.
Error bars are less than the symbol size.
Lines are guides to the eye.
}
\label{f5}
\end{figure}

The elastic energy $E_{\rm el}(A)$ is defined here as the increase 
in energy corresponding to a strictly planar graphene layer with area
$A$ respect to the minimum energy $E_0$.
Thus, we have calculated $E_{\rm el}(A)$ for a supercell including 
960 carbon atoms, expanding it isotropically and keeping it flat.
Finite-size effects on the elastic energy are very small, and in fact 
negligible for our current purposes, as happens for the size dependence
of the area $A$.
For $A > A_0$, the elastic energy increases with $A$, and for small
lattice expansion it can be approximated as
$E_{\rm el}(A) \approx C (A - A_0)^2$, with $C$ = 2.41 eV \AA$^{-2}$.
At room temperature ($T \sim$ 300~K) and for small stresses $P$
($A$ close to $A_0$), the elastic energy is much smaller than 
the vibrational energy $E_{\rm vib}$, but this can be different for
low $T$ and/or large applied stresses (see below).
Once calculated the elastic energy for the area $A$ resulting from
the simulations at given $T$ and $P$,
we obtain the vibrational energy $E_{\rm vib}(A,T)$ by subtracting 
the elastic energy from the internal energy $E(T)$:
$E_{\rm vib} = E(T) - E_0 - E_{\rm el}(A)$ (see Eq.~(\ref{et})).
Then, the potential energy corresponding to vibrational motion,
$V_{\rm vib}$, is found as $V_{\rm vib} = E_{\rm vib} - K$.

In Fig.~5 we present the different contributions to the internal energy 
of graphene vs. tensile stress, as derived from PIMD simulations for
$N$ = 960 atoms and $T$ = 25~K.
In this figure, diamonds represent the elastic energy, and circles and
squares indicate $K$ and $V_{\rm vib}$, respectively.
At $P = 0$, $E_{\rm el}$ is close to zero, but slightly positive, as a
consequence of the zero-point expansion of the graphene lattice, which
causes that $A > A_0$.  For increasing tensile stress, $E_{\rm el}$ 
rises and becomes similar to $K$ and $V_{\rm vib}$ for 
$P \approx -0.6$ eV \AA$^{-2}$. At still larger tensile stress, 
the elastic contribution is the largest one, as shown in Fig.~5.
For $T$ = 300 K the picture is qualitatively the same. The elastic
energy increases roughly a constant value (26 meV/atom) with respect to 
the results at 25 K in the whole stress range shown in Fig.~5.
The same happens for the kinetic energy, with a rise of 6 meV/atom.
As a result, the crossing of $E_{\rm el}$ and $K$ at 300 K occurs
for a tensile stress $P \approx -0.55$ eV \AA$^{-2}$.

For a purely harmonic model for the vibrational modes, one expects
$K = V_{\rm vib}$ (virial theorem\cite{la80,fe72}), i.e., an energy 
ratio $K/V_{\rm vib} = 1$
at any temperature in both classical and quantum approaches.
This is not strictly the case for the results of our PIMD, because
the vibrational amplitudes are finite, even at low temperatures, and feel 
the anharmonicity of the interatomic potential.
In particular, we find $K > V_{\rm vib}$, for all temperatures and 
tensile stresses considered here.  
As displayed in Fig.~5 for $T$ = 25~K, 
the difference $K - V_{\rm vib}$ increases as the tensile stress is raised,
so that $K$ is about 5\% larger than $V_{\rm vib}$ for small stress, and 
around 9\% larger for a stress of $-0.75$ eV \AA$^{-2}$.
Differences between the kinetic and potential contribution to the
vibrational energy have been used for a quantification of the
anharmonicity in condensed matter, as discussed earlier from
path-integral simulations, e.g., for van der Waals solids\cite{he02}
and H impurities in silicon.\cite{he95}

Concerning the energy results for our quantum approach at low $T$,
we note that analyses of anharmonicity in solids, based on 
quasiharmonic approximations and perturbation theory indicate 
that low-temperature changes in the vibrational energy with respect to 
a harmonic calculation are mostly due to the kinetic energy.
Thus, considering perturbed harmonic oscillators with perturbations
of type $x^3$ or $x^4$ at $T = 0$, first-order changes in the
energy are due to changes in $K$, and the potential
energy stays unshifted in its unperturbed value.\cite{la65,he95}

\section{Out-of-plane motion}

In this section we study the mean-square displacements
of carbon atoms in the $z$ direction, normal to the graphene sheet, 
as obtained from our PIMD simulations.
We mostly concentrate on the nature of these atomic displacements, 
i.e., if they can be well described by a classical model, or the C atoms 
appreciably behave as quantum particles. We expect of course that 
a classical description will lose accuracy as the temperature
is reduced, but in the case of graphene it has been shown earlier that 
other factors such as the system size play also an important role
in this question.\cite{he16} Moreover, an external stress modifies 
the vibrational
frequencies in the material, thus causing a change in the vibrational
amplitudes and in the accuracy of a classical description at a given 
temperature. 

PIMD simulations can be used to study vibrational amplitudes or atomic
delocalization at finite temperatures.
This includes a thermal (classical-like) motion,
as well as a delocalization due to the quantum nature of the atomic
nuclei, which can be quantified by the spacial extension of the paths
associated to a given atomic nucleus.
For each quantum path, we define the center-of-gravity (centroid) as
\begin{equation}
   \overline{\bf r}_i = \frac{1}{N_{\rm Tr}} 
          \sum_{j=1}^{N_{\rm Tr}} {\bf r}_{ij} \; ,
\label{centr}
\end{equation}
where ${\bf r}_{ij} \equiv (x_{ij}, y_{ij}, z_{ij})$ is the 3D position 
of bead $j$ in the ring polymer associated to nucleus $i$. 
For the out-of-plane motion, 
we focus on the $z$-coordinate of the polymer beads.
Then, the mean-square displacement $(\Delta z)^2_i$ of the atomic nucleus
$i$ in the $z$ direction along a PIMD simulation run is defined as
\begin{equation}
  (\Delta z)^2_i =  \frac{1}{N_{\rm Tr}} \left< \sum_{j=1}^{N_{\rm Tr}}
           ( z_{ij} - \left< \overline{z}_i \right>)^2
           \right>    \, ,
\label{deltaz2}
\end{equation}

The kinetic energy of a particle is related to its quantum delocalization,
or in the present context, to the spread of the paths associated to it.
This can be measured by the mean-square {\em radius-of-gyration} of
the ring polymers, with an out-of-plane component:\cite{gi88,gi90}
\begin{equation}
  Q_{z,i}^2 = \frac{1}{N_{\rm Tr}} \left< \sum_{j=1}^{N_{\rm Tr}}
             (z_{ij} - \overline{z}_i)^2 \right>    \, .
\label{qz2}
\end{equation}
The total spatial delocalization $(\Delta z)^2_i$ of atomic nucleus $i$ in 
the $z$ direction at a finite temperature includes, in addition to 
$Q_{z,i}^2$, another contribution which accounts for the classical-like 
motion of the centroid coordinate $\overline{z}_i$, i.e.
\begin{equation}
    (\Delta z)^2_i = Q_{z,i}^2 + C_{z,i}^2  \, ,
\label{deltaz2b}
\end{equation}
with
\begin{equation}
 C_{z,i}^2 = \left< \left( \overline{z}_i - \langle \overline{z}_i \rangle
                \right)^2 \right>
       =  \langle  \overline{z}_i^2 \rangle -
          \langle  \overline{z}_i \rangle^2  \, .
\label{cz2}
\end{equation}
This term $C_{z,i}^2$ converges at high $T$ to the mean-square displacement
$(\Delta z)^2_{i,{\rm cl}}$ given by a classical model, since in this limit
each quantum path collapses onto a single point ($Q_{z,i}^2 \to 0$).
In the results presented below, we will show data for 
$(\Delta z)^2$, $Q_z^2$, and $C_z^2$, calculated as averages for the
$N$ atoms in a simulation cell. For example:
\begin{equation}
    (\Delta z)^2 = \frac1N  \sum_{i=1}^N  (\Delta z)^2_i   \, .
\end{equation} 
 
To connect the results of our simulations with the 
out-of-plane displacements corresponding to vibrational modes of the
graphene sheet, we recall that the atomic mean-square displacement 
at temperature $T$ is given in a HA by
\begin{equation}
 (\Delta z)^2_{\rm HA} = \frac1N \sum_{i,\bf k} 
           \frac{\hbar}{2 m \omega_i({\bf k})}
       \coth \left( \frac{\hbar \omega_i({\bf k})}{2 k_B T}  \right)  \, ,
\label{qz2b}
\end{equation}
where the index $i$ ($i$ = 1, 2) refers to the phonon bands ZA and ZO,
with atomic displacements along the $z$ direction.\cite{mo05,ka11,wi04}
The sum in ${\bf k}$ is extended to wavevectors
${\bf k} = (k_x, k_y)$ in the hexagonal Brillouin zone,
with discrete ${\bf k}$ points spaced by $\Delta k_x = 2 \pi / L_x$ and
$\Delta k_y = 2 \pi / L_y$.\cite{ra16}
Eq.~(\ref{qz2b}) has been used here to calculate $(\Delta z)^2_{\rm HA}$ 
in a harmonic approach.
Increasing the system size $N$ causes the appearance of vibrational modes
with longer wavelength $\lambda$. In fact, one has for the phonons
an effective wavelength cut-off
$\lambda_{max} \approx L$, with $L = (N A_p)^{1/2}$,
and the minimum wavevector is
$k_0 = 2 \pi / \lambda_{max}$, i.e., $k_0 \sim N^{-1/2}$.
For the calculations presented below, based on the formula in Eq.~(\ref{qz2b}),
we have used the vibrational frequencies in the phonon branches ZA and ZO
obtained from diagonalization of the dynamical matrix corresponding to the
LCBOPII potential employed here.

\begin{figure}
\vspace{-1.0cm}
\includegraphics[width=8.5cm]{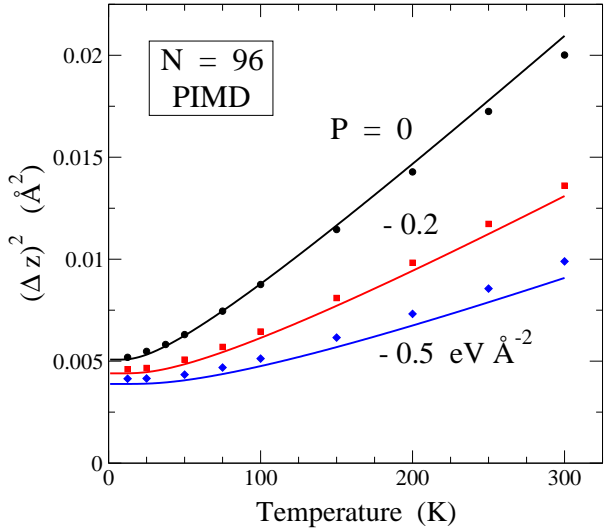}
\vspace{-0.5cm}
\caption{Mean-square displacement in the $z$ direction as a function
of temperature, as derived from PIMD simulations for $N$ = 96 and
$P$ = 0 (circles), $-0.2$ (squares), and $-0.5$ eV \AA$^{-2}$ (diamonds).
Error bars are in the order of the symbol size.
Bold lines correspond to a harmonic approximation based on ZA
and ZO vibrational modes.
}
\label{f6}
\end{figure}

For an applied stress $P$, the most important effect in the ZA and ZO
bands is a change in frequency of the ZA modes in the low-frequency
region, for which
\begin{equation}  
  \omega_{\rm ZA}({\bf k})^2 = \omega^0_{\rm ZA}({\bf k})^2 - 
          \frac{P}{\rho} \, k^2
\label{omega2}
\end{equation}  
The zero-stress band $\omega^0_{\rm ZA}({\bf k})$ calculated for the
minimum-energy structure (area $A_0$), verifies for small $k$:
$\rho \, \omega^0_{\rm ZA}({\bf k})^2 \approx \kappa k^4$.
Then, for $P < 0$ the small-$k$ region is dominated by the quadratic term
(linear in $P$) in Eq.~(\ref{omega2}), so that 
$\omega_{\rm ZA}({\bf k}) \approx \sqrt{-P/\rho} \; k$
for $k \ll$ 1~\AA$^{-1}$.

In Fig.~6 we show results for the motion in the out-of-plane
direction, obtained for a cell including 96 atoms. 
The use of a relatively small simulation cell is convenient to visualize
the behavior of $(\Delta z)^2$ in the low-temperature region, where 
quantum effects are prominent. For larger cell sizes, these effects appear
only at lower temperatures, which turns out to be difficult to observe
from PIMD simulations.\cite{he16} 
Solid symbols represent the mean-square displacement $(\Delta z)^2$ 
derived from the simulations, as a function of temperature for three 
different applied stresses: 
$P = 0$ (circles), $-0.2$ (squares), and $-0.5$ eV \AA$^{-2}$ (diamonds).
Lines were calculated from a harmonic approximation based on
the ZA and ZO phonon bands of graphene, as indicated above.

\begin{figure}
\vspace{-1.0cm}
\includegraphics[width=8.5cm]{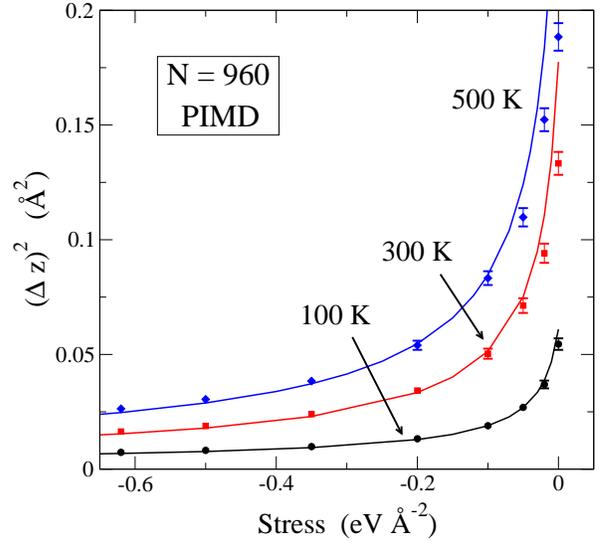}
\vspace{-0.5cm}
\caption{Mean-square displacement in the $z$ direction as a function
of tensile stress, as derived from PIMD simulations for $N$ = 960 and
three temperatures:
$T$ = 500 K (diamonds), 300 K (squares), and 100 K (circles).
Error bars, when not displayed, are in the order or less  than the symbol
size.  Bold lines correspond to a harmonic approximation based on ZA
and ZO vibrational modes.
}
\label{f7}
\end{figure}

One observes in Fig.~6 that the vibrational amplitude decreases as the 
tensile stress increases, mainly due to an increase in vibrational 
frequencies of ZA modes with low $k$ ($k \ll$ 1 \AA$^{-1}$).
This becomes important as the temperature is raised, but it is also 
appreciable in the low-temperature region, as shown in the figure.
The lines derived from a HA are close to the results of the 
PIMD simulations at low temperature, but both sets of results depart
progressively one from the other as temperature is raised.
For $P = 0$, $(\Delta z)^2_{\rm HA} < (\Delta z)^2_{\rm PI}$, 
but the opposite happens for $T >$ 100~K.
For relatively large tensile stresses of $-0.2$ and $-0.5$ eV/\AA$^{-2}$,
we find $(\Delta z)^2_{\rm HA} < (\Delta z)^2_{\rm PI}$ in the whole
temperature range presented in Fig.~6. For $T > 300$~K, the
difference between $(\Delta z)^2_{\rm PI}$ and $(\Delta z)^2_{\rm HA}$
steadily increases.

In Fig.~7 we display the mean-square displacements $(\Delta z)^2$ 
for $N = 960$ as a function of applied stress $P$ for three temperatures: 
100, 300, and 500~K. Symbols are results of PIMD simulations, whereas the
lines correspond to the HA based on ZA and ZO vibrational modes. 
One observes first an important decrease in $(\Delta z)^2$
as the tensile stress is raised. This decrease is most appreciable
for stresses in the range from 0 to $-0.1$ eV \AA$^{-2}$. For larger
stresses, the reduction of $(\Delta z)^2$ becomes slower.
One also notices that the largest difference between 
$(\Delta z)^2_{\rm PI}$ and $(\Delta z)^2_{\rm HA}$ occurs for $P = 0$,
and it becomes smaller for larger pressure. For $P \sim -0.2$ eV \AA$^{-2}$,
both sets of results cross each other and the data derived from PIMD become
slightly larger than those corresponding to the HA.

\begin{figure}
\vspace{-1.0cm}
\includegraphics[width=8.5cm]{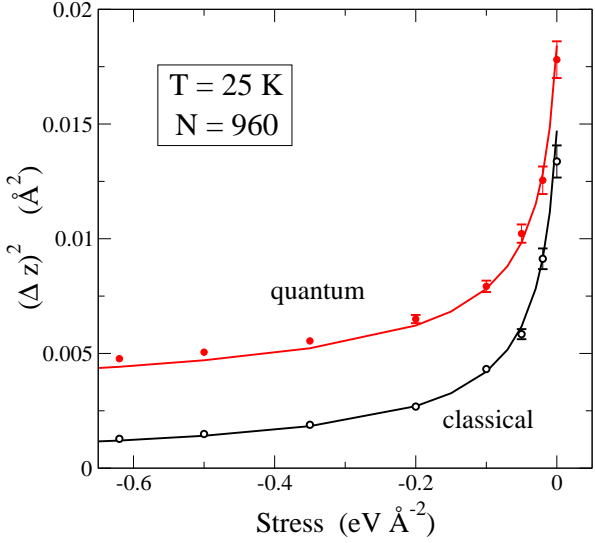}
\vspace{-0.5cm}
\caption{Mean-square displacement in the $z$ direction as a function
of tensile stress, as derived from PIMD (solid circles) and classical
(open circles) simulations for $N$ = 960 and $T$ = 25~K.
Error bars, when not shown, are in the order or less than the symbol
size.  Lines were obtained from classical and quantum harmonic
approximations based on ZA and ZO vibrational modes.
}
\label{f8}
\end{figure}

To get better insight into the influence of nuclear quantum effects on
$(\Delta z)^2$ for different pressures, we have plotted in Fig.~8 the 
mean-square displacements as derived from classical MD (open circles) and
PIMD (solid circles) simulations at 25~K.
The lines were obtained from the HA described above for the quantum case
[see Eq.~(\ref{qz2b})],
and for the classical calculation we used the expression
\begin{equation}
    C_{z,\rm HA}^2  = \frac1N \sum_{i,\bf k}
               \frac{k_B T}{m \omega_i({\bf k})^2}    \, .
\label{cz2b}
\end{equation}
One sees that the relevance of quantum effects increases for rising
tensile stress, as can be measured from the ratio
$(\Delta z)^2 / C_z^2$ between the mean-square displacements in the
quantum and classical case. In fact, this ratio goes from 1.3 for
$P = 0$ to 3.7 for $P = -0.6$ eV \AA$^{-2}$. 
For a given stress, the difference between the quantum results for 
$(\Delta z)^2$ and the classical ones, $C_z^2$, decreases as 
temperature is raised. In fact, at $T = 300$ K the mean-square 
displacement derived from PIMD simulations is about 10\% larger than 
the classical result in the stress range displayed in Fig.~8. 

\begin{figure}
\vspace{-1.0cm}
\includegraphics[width=8.5cm]{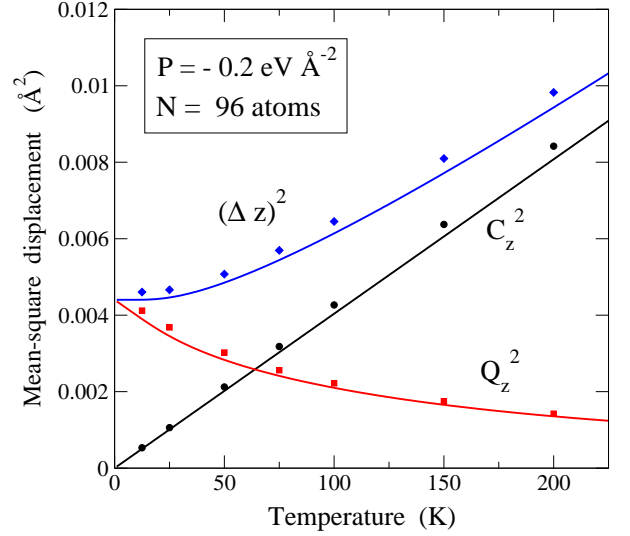}
\vspace{-0.5cm}
\caption{
Temperature dependence of the mean-square displacement along
the out-of-plane direction, $(\Delta z)^2$ (diamonds),
along with its classical $C_z^2$ (circles) and quantum
$Q_z^2$ (squares) contributions.
These data correspond to PIMD simulations for a graphene cell
containing 96 atoms and a tensile stress $P = -0.2$~eV \AA$^{-2}$.
Lines were obtained from a harmonic approximation based on ZA and ZO
vibrational modes.
}
\label{f9}
\end{figure}

As indicated above, the mean-square displacement $(\Delta z)^2$ can
be divided into two parts, $Q_z^2$ and $C_z^2$, the first one properly 
quantum in nature, measuring the extension of the quantum paths,
and the second of a classical-like character, taking account of
the centroid motion, i.e., global displacements of the paths.
For the sake of comparing with the results of PIMD simulations, 
we have also calculated in the harmonic approximation 
$Q_{z,\rm HA}^2 = (\Delta z)^2_{\rm HA} - C_{z,\rm HA}^2$, using
Eqs.~(\ref{qz2b}) and (\ref{cz2b}).
To visualize the evolution of both terms as a function of
temperature, we have plotted in Fig.~9 $(\Delta z)^2$ along with its
two contributions $Q_z^2$ and  $C_z^2$
for $N$ = 96 and a stress $P = -0.2$~eV \AA$^{-2}$.  
In the limit $T \to 0$, $C_z^2$ vanishes and $Q_z^2$ converges to a value 
of about $4.5 \times 10^{-3}$ \AA$^2$. 
$Q_z^2$ decreases for increasing temperature, as nuclear quantum effects 
become less relevant. On the contrary, the classical contribution $C_z^2$ 
increases with $T$, linearly in the HA [see Eq.~(\ref{cz2b})], and almost
linearly for the results of PIMD simulations.

For the system size shown in Fig.~9 (N = 96), 
both contributions to $(\Delta z)^2$ are equal at $T \approx$ 64~K. 
At higher temperatures, the classical-like part $C_z^2$ is the main 
contribution to the atomic displacements in the $z$ direction.
Values of $C_z^2$ given by PIMD simulations coincide within error
bars with the mean-square atomic displacement obtained from classical 
MD simulations.
The actual quantum delocalization can be estimated from the mean extension 
of the quantum paths in the $z$ direction, i.e., from $Q_z^2$. 
For $P = -0.2$~eV \AA$^{-2}$, we find at 25 and 300~K an average 
extension $(\Delta z)_Q = (Q_z^2)^{1/2} \approx$ 0.06 and 0.03 \AA, 
respectively. 

\begin{figure}
\vspace{-1.0cm}
\includegraphics[width=8.5cm]{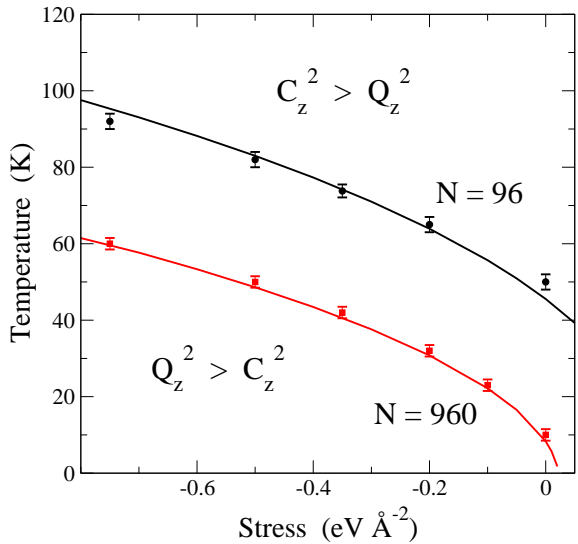}
\vspace{-0.5cm}
\caption{$P - T$ plane displaying the crossover from the region dominated by
quantum delocalization ($Q_z^2 > C_z^2$, below the lines) to
the region dominated by classical-like motion ($C_z^2 > Q_z^2$,
above the lines).
Data points were obtained from PIMD simulations for two system sizes:
$N$ = 96 (circles) and 960 (squares).
The lines were obtained from the HA discussed in the text.
}
\label{f10}
\end{figure}

The picture displayed in Fig.~9 for the atom displacements in the 
out-of-plane direction is qualitatively the same for different system 
sizes and applied stresses, but the temperature region where 
$Q_z^2$ or $C_z^2$ is the main contribution to $(\Delta z)^2$ as well as
the crossing point $Q_z^2 = C_z^2$ greatly depend on both 
variables $N$ and $P$.
The dependence on $N$ is mainly caused by the enhancement of the 
classical-like contribution $C_z^2$ for increasing size, as
observed earlier for results of classical 
MD simulations.\cite{lo09,ga14,ra16}
The quantum contribution $Q_z^2$ has a small size effect and 
converges fast as $N$ is increased.\cite{he16}
For a given system size $N$, the ratio $Q_z^2 / C_z^2$ decreases for 
increasing $T$ (see Fig.~9), and there is a crossover temperature $T_c(P)$ 
for which this ratio is unity. For $T > T_c(P)$ classical-like motion 
dominates the atomic motion in the $z$ direction.

In Fig.~10 we present $T_c$ as a function of the applied stress $P$ for 
two system sizes: $N$ = 96 and 960.
Symbols are results derived from PIMD simulations and
lines correspond to crossover temperatures derived from the HA based
on the ZA and ZO phonon bands.
Below the lines we have $Q_z^2 > C_z^2$, i.e. the quantum contribution 
dominates in $(\Delta z)^2$, and the opposite happens above the lines
with classical-like motion dominating the out-of-plane displacements.
The crossover temperature $T_c$ depends on the system size $N$,
since the effective low-energy cutoff scales as $k_0 \sim N^{-1/2}$.
This means that for a given stress $P$, an increase in $N$ causes
the appearance of ZA modes with lower frequencies, which contribute to
reduce $T_c$ (for a given $\omega$, the classical behavior dominates for
$T \gtrsim \hbar \omega / k_B$). 

The main limitations of the HA are the neglect of anharmonicity in 
the transverse vibrational modes, which is expected to be reasonably 
small at low $T$, and the use of Eq.~(\ref{omega2}) for ZA mode
frequencies under stress. This equation, giving $\omega_{\rm ZA}({\bf k})^2$ 
as a sum of a zero-stress term, $\omega^0_{\rm ZA}({\bf k})^2$, 
and a term linear in the applied stress, $P k^2 / \rho$, is expected to 
be valid for small-$k$ ZA modes, which dominate the mean-square
displacement of C atoms in the transverse direction to the graphene plane. 
Taking into account these limitations, the harmonic model captures 
qualitatively, and almost quantitatively, the basic aspects 
of the competition between classical-like and quantum dynamics
of the C atoms in the $z$ direction.

\section{Summary}

We have presented results of PIMD simulations of a graphene
monolayer at several temperatures and tensile stresses.
The importance of quantum effects has been quantified by comparing results
of such quantum simulations with those obtained from classical MD simulations.
Structural variables are found to change when quantum nuclear motion is
taken into account, especially at low temperatures.
Thus, the sheet area and interatomic distances change appreciably
in the range of temperatures and stresses considered here.

The LCBOPII potential model was shown earlier to give a
reliable description of structural and thermodynamic properties 
of graphene.  We have investigated here its reliability to describe 
nuclear quantum effects in the presence of a tensile stress. 
The results obtained in the simulations have allowed us to analyze
both the in-plane $A_p$ and real area $A$ of graphene as functions
of $T$ and $P$. 
The difference $A - A_p$ grows (decreases) as temperature (stress) 
is raised.
The thermal contraction of $A_p$ becomes less important as the tensile
stress increases and the amplitude of out-of-plane vibrations decreases.
We emphasize that PIMD simulations yield a negative thermal expansion 
of $A_p$ at low $T$ for pressures so high as $-0.5$ eV \AA$^{-2}$
($-8$~N/m). On the contrary, in classical calculations this thermal 
expansion becomes positive at much smaller stresses.

 Zero-point expansion of the graphene layer due to nuclear quantum
effects is not negligible, and it amounts to an increase of about 1\% in
the area $A$. This zero-point effect is reduced for rising tensile stress.
Moreover, the temperature dependence of the in-plane area $A_p$ is
qualitatively different when derived from PIMD or classical simulations,
even at temperatures between 300 and 1000 K.
Such a difference appears for all tension stresses considered here, up to
$P = -0.5$ eV \AA$^{-2}$.

Atomic vibrations in the out-of-plane $z$ direction have been
particularly considered, as they are important for the area $A_p$ and
for the relative stability of the planar graphene layer vs. crumpling.
Quantum effects are dominant for these vibrational modes provided that
$k_B T < \hbar \omega$. However, the actual atomic motion at any finite 
temperature, resulting from the sum of mode contributions, is dominated
by the classical thermal contribution as soon as the system size is 
large enough. This size effect appears in the quantum simulations at low 
temperatures, as a result of the presence of vibrational modes in the 
ZA band with smaller wavenumbers (frequencies) for larger graphene cells.
In this respect, an interesting result is that the temperature region
where quantum motion is dominant is enhanced by an external tensile stress,
as shown in Fig.~10, i.e., for a given system size $T_c$ increases 
as the stress is raised.

An important point related to the consistency of the simulation
results is their agreement with the principles of thermodynamics, in
particular with the third law. 
This means that thermal expansion coefficients should converge to zero 
for $T \to 0$.
We have found that this requirement is verified by both the in-plane 
area $A_p$ and the real area $A$ obtained from PIMD simulations
for $P = 0$ and $P < 0$ (tensile stress).

An analysis of graphene under tensile stresses larger than those 
considered here may be interesting for its effect on mechanical,
electronic, and optical properties. Such a study can be hindered
by limitations associated to effective potentials at large strains,
and would require the use of {\em ab-initio} methods (e.g., 
density-functional theory).
An efficient combination of these methods with path-integral 
simulations is still restricted to cell sizes relatively small,
a limitation that is expected to be progressively overcome 
in the forthcoming years.

\begin{acknowledgments}
The authors acknowledge the support of J. H. Los in the implementation 
of the LCBOPII potential.
This work was supported by Direcci\'on General de Investigaci\'on,
MINECO (Spain) through Grants FIS2012-31713 and FIS2015-64222-C2.
\end{acknowledgments}


\vspace{1cm}

\appendix

\section{Pressure estimator}
\label{app1}

The Cartesian coordinates of the $N$ atomic nuclei in the crystalline
membrane are denoted as ${\bf r}_{ij}$, where the $i = 1, ..., N$ indicates
the atom, and $j = 1, ..., N_{\rm Tr}$ refers to the bead number.
The staging coordinates ${\bf u}_{ij}$ are defined by a linear
transformation of ${\bf r}_{ij}$ that diagonalizes 
the {\em harmonic} energy associated to the effective interactions 
between neighboring beads:\cite{tu98,tu02}
\begin{equation}
  {\bf u}_{i1} = {\bf r}_{i1} \; ,
\end{equation}
\begin{equation}
  {\bf u}_{ij} = {\bf r}_{ij} - \frac{j-1}{j} {\bf r}_{i,j+1} - 
         \frac{1}{j} {\bf r}_{i1} \; ,   \;\; j = 2, \dots, N_{\rm Tr} \; .
\end{equation}

The estimator of the 2D stress tensor ${\bf \tau}$ is
similar to that employed in previous work for 3D systems.\citep{ra08,he14}
For PIMD simulations, its components are given by expressions
such as\cite{tu98}
\begin{widetext}
\begin{equation}
 \tau_{xy} = \left \langle \frac{1}{A_p}
    \left(  \sum_{i=1}^{N}  \sum_{j=1}^{N_{\rm Tr}}
       \left( m_j v_{ij,x} v_{ij,y} - 2 k_j u_{ij,x} u_{ij,y}  \right) -
     \frac{1}{N_{\rm Tr}} \sum_{j=1}^{N_{\rm Tr}} 
     \frac {\partial U({\bf r}_{1j}, \dots, {\bf r}_{Nj})}
           {\partial \epsilon_{xy}} \right) \right \rangle  \; ,
\end{equation}
\end{widetext}
where the brackets  $\langle \cdots \rangle$ indicate an ensemble average,
$m_j$ is the dynamic mass associated to the staging coordinate
${\bf u}_{ij}$, $v_{ij,x}$, $v_{ij,y}$ are components of its
corresponding velocity, and
$\epsilon_{xy}$ is an element of the 2D strain tensor.
The masses $m_j$ are given by\cite{tu98,tu02,he14}
\begin{equation}
  m_1 = m   \; ,
\end{equation}
\begin{equation}
  m_j = \frac{j}{j-1} m   \; ,  \;\; j = 2 , \dots, N_{\rm Tr} \; ,
\end{equation}
where $m$ is the nuclear mass.
The constant $k_j$ is given for each bead $j$ by
$k_j = m_j N_{\rm Tr} / 2 \beta^2 \hbar^2$ for $j > 1$ and $k_1 = 0$
($\beta = (k_B T)^{-1}$).

The stress $P$ used in our PIMD simulations is 
\begin{equation}
  P = \frac{1}{2} \left( \tau_{xx} + \tau_{yy} \right)  \, ,
\end{equation}
or $P = \langle {\cal P} \rangle$ with the estimator 
\begin{multline}
  {\cal P} = \frac{1}{2 A_p} \sum_{i=1}^N  \sum_{j=1}^{N_{\rm Tr}}
           m_j \left( v_{ij,x}^2 + v_{ij,y}^2 \right) 
        - \frac{E_h}{A_p}  \\
        - \frac{1}{N_{\rm Tr}} \sum_{j=1}^{N_{\rm Tr}}
        \frac{\partial U({\bf r}_{1j}, \dots, {\bf r}_{Nj})} {\partial A_p}
            \; .
\label{calP}
\end{multline}
$E_h$ is the in-plane {\em harmonic} energy, that in terms of staging 
coordinates can be written as
\begin{equation}
  E_h =  \sum_{i=1}^N  \sum_{j=2}^{N_{\rm Tr}}
           k_j \left( u_{ij,x}^2 + u_{ij,y}^2 \right)   \; .
\end{equation}
The last term in Eq.~(\ref{calP}) is a sum of derivatives of the potential
energy $U$ at different imaginary times (beads) $j$.
In the present case of graphene, these derivatives were carried out 
analytically.



\begin{thebibliography}{84}
\expandafter\ifx\csname natexlab\endcsname\relax\def\natexlab#1{#1}\fi
\expandafter\ifx\csname bibnamefont\endcsname\relax
  \def\bibnamefont#1{#1}\fi
\expandafter\ifx\csname bibfnamefont\endcsname\relax
  \def\bibfnamefont#1{#1}\fi
\expandafter\ifx\csname citenamefont\endcsname\relax
  \def\citenamefont#1{#1}\fi
\expandafter\ifx\csname url\endcsname\relax
  \def\url#1{\texttt{#1}}\fi
\expandafter\ifx\csname urlprefix\endcsname\relax\def\urlprefix{URL }\fi
\providecommand{\bibinfo}[2]{#2}
\providecommand{\eprint}[2][]{\url{#2}}

\bibitem[{\citenamefont{Geim and Novoselov}(2007)}]{ge07}
\bibinfo{author}{\bibfnamefont{A.~K.} \bibnamefont{Geim}} \bibnamefont{and}
  \bibinfo{author}{\bibfnamefont{K.~S.} \bibnamefont{Novoselov}},
  \bibinfo{journal}{Nature Mater.} \textbf{\bibinfo{volume}{6}},
  \bibinfo{pages}{183} (\bibinfo{year}{2007}).

\bibitem[{\citenamefont{Castro~Neto et~al.}(2009)\citenamefont{Castro~Neto,
  Guinea, Peres, Novoselov, and Geim}}]{ca09b}
\bibinfo{author}{\bibfnamefont{A.~H.} \bibnamefont{Castro~Neto}},
  \bibinfo{author}{\bibfnamefont{F.}~\bibnamefont{Guinea}},
  \bibinfo{author}{\bibfnamefont{N.~M.~R.} \bibnamefont{Peres}},
  \bibinfo{author}{\bibfnamefont{K.~S.} \bibnamefont{Novoselov}},
  \bibnamefont{and} \bibinfo{author}{\bibfnamefont{A.~K.} \bibnamefont{Geim}},
  \bibinfo{journal}{Rev. Mod. Phys.} \textbf{\bibinfo{volume}{81}},
  \bibinfo{pages}{109} (\bibinfo{year}{2009}).

\bibitem[{\citenamefont{Flynn}(2011)}]{fl11}
\bibinfo{author}{\bibfnamefont{G.~W.} \bibnamefont{Flynn}},
  \bibinfo{journal}{J. Chem. Phys.} \textbf{\bibinfo{volume}{135}},
  \bibinfo{pages}{050901} (\bibinfo{year}{2011}).

\bibitem[{\citenamefont{Roldan et~al.}(2017)\citenamefont{Roldan, Chirolli,
  Prada, Angel Silva-Guillen, San-Jose, and Guinea}}]{ro17}
\bibinfo{author}{\bibfnamefont{R.}~\bibnamefont{Roldan}},
  \bibinfo{author}{\bibfnamefont{L.}~\bibnamefont{Chirolli}},
  \bibinfo{author}{\bibfnamefont{E.}~\bibnamefont{Prada}},
  \bibinfo{author}{\bibfnamefont{J.}~\bibnamefont{Angel Silva-Guillen}},
  \bibinfo{author}{\bibfnamefont{P.}~\bibnamefont{San-Jose}}, \bibnamefont{and}
  \bibinfo{author}{\bibfnamefont{F.}~\bibnamefont{Guinea}},
  \bibinfo{journal}{Chem. Soc. Rev.} \textbf{\bibinfo{volume}{46}},
  \bibinfo{pages}{4387} (\bibinfo{year}{2017}).

\bibitem[{\citenamefont{Ghosh et~al.}(2008)\citenamefont{Ghosh, Calizo,
  Teweldebrhan, Pokatilov, Nika, Balandin, Bao, Miao, and Lau}}]{gh08b}
\bibinfo{author}{\bibfnamefont{S.}~\bibnamefont{Ghosh}},
  \bibinfo{author}{\bibfnamefont{I.}~\bibnamefont{Calizo}},
  \bibinfo{author}{\bibfnamefont{D.}~\bibnamefont{Teweldebrhan}},
  \bibinfo{author}{\bibfnamefont{E.~P.} \bibnamefont{Pokatilov}},
  \bibinfo{author}{\bibfnamefont{D.~L.} \bibnamefont{Nika}},
  \bibinfo{author}{\bibfnamefont{A.~A.} \bibnamefont{Balandin}},
  \bibinfo{author}{\bibfnamefont{W.}~\bibnamefont{Bao}},
  \bibinfo{author}{\bibfnamefont{F.}~\bibnamefont{Miao}}, \bibnamefont{and}
  \bibinfo{author}{\bibfnamefont{C.~N.} \bibnamefont{Lau}},
  \bibinfo{journal}{Appl. Phys. Lett.} \textbf{\bibinfo{volume}{92}},
  \bibinfo{pages}{151911} (\bibinfo{year}{2008}).

\bibitem[{\citenamefont{Nika et~al.}(2009)\citenamefont{Nika, Pokatilov,
  Askerov, and Balandin}}]{ni09}
\bibinfo{author}{\bibfnamefont{D.~L.} \bibnamefont{Nika}},
  \bibinfo{author}{\bibfnamefont{E.~P.} \bibnamefont{Pokatilov}},
  \bibinfo{author}{\bibfnamefont{A.~S.} \bibnamefont{Askerov}},
  \bibnamefont{and} \bibinfo{author}{\bibfnamefont{A.~A.}
  \bibnamefont{Balandin}}, \bibinfo{journal}{Phys. Rev. B}
  \textbf{\bibinfo{volume}{79}}, \bibinfo{pages}{155413}
  (\bibinfo{year}{2009}).

\bibitem[{\citenamefont{Balandin}(2011)}]{ba11}
\bibinfo{author}{\bibfnamefont{A.~A.} \bibnamefont{Balandin}},
  \bibinfo{journal}{Nature Mater.} \textbf{\bibinfo{volume}{10}},
  \bibinfo{pages}{569} (\bibinfo{year}{2011}).

\bibitem[{\citenamefont{Lee et~al.}(2008)\citenamefont{Lee, Wei, Kysar, and
  Hone}}]{le08}
\bibinfo{author}{\bibfnamefont{C.}~\bibnamefont{Lee}},
  \bibinfo{author}{\bibfnamefont{X.}~\bibnamefont{Wei}},
  \bibinfo{author}{\bibfnamefont{J.~W.} \bibnamefont{Kysar}}, \bibnamefont{and}
  \bibinfo{author}{\bibfnamefont{J.}~\bibnamefont{Hone}},
  \bibinfo{journal}{Science} \textbf{\bibinfo{volume}{321}},
  \bibinfo{pages}{385} (\bibinfo{year}{2008}).

\bibitem[{\citenamefont{Seol et~al.}(2010)\citenamefont{Seol, Jo, Moore,
  Lindsay, Aitken, Pettes, Li, Yao, Huang, Broido et~al.}}]{se10}
\bibinfo{author}{\bibfnamefont{J.~H.} \bibnamefont{Seol}},
  \bibinfo{author}{\bibfnamefont{I.}~\bibnamefont{Jo}},
  \bibinfo{author}{\bibfnamefont{A.~L.} \bibnamefont{Moore}},
  \bibinfo{author}{\bibfnamefont{L.}~\bibnamefont{Lindsay}},
  \bibinfo{author}{\bibfnamefont{Z.~H.} \bibnamefont{Aitken}},
  \bibinfo{author}{\bibfnamefont{M.~T.} \bibnamefont{Pettes}},
  \bibinfo{author}{\bibfnamefont{X.}~\bibnamefont{Li}},
  \bibinfo{author}{\bibfnamefont{Z.}~\bibnamefont{Yao}},
  \bibinfo{author}{\bibfnamefont{R.}~\bibnamefont{Huang}},
  \bibinfo{author}{\bibfnamefont{D.}~\bibnamefont{Broido}},
  \bibnamefont{et~al.}, \bibinfo{journal}{Science}
  \textbf{\bibinfo{volume}{328}}, \bibinfo{pages}{213} (\bibinfo{year}{2010}).

\bibitem[{\citenamefont{Prasher}(2010)}]{pr10}
\bibinfo{author}{\bibfnamefont{R.}~\bibnamefont{Prasher}},
  \bibinfo{journal}{Science} \textbf{\bibinfo{volume}{328}},
  \bibinfo{pages}{185} (\bibinfo{year}{2010}).

\bibitem[{\citenamefont{Herrero and Ram\'irez}(2016)}]{he16}
\bibinfo{author}{\bibfnamefont{C.~P.} \bibnamefont{Herrero}} \bibnamefont{and}
  \bibinfo{author}{\bibfnamefont{R.}~\bibnamefont{Ram\'irez}},
  \bibinfo{journal}{J. Chem. Phys.} \textbf{\bibinfo{volume}{145}},
  \bibinfo{pages}{224701} (\bibinfo{year}{2016}).

\bibitem[{\citenamefont{Fasolino et~al.}(2007)\citenamefont{Fasolino, Los, and
  Katsnelson}}]{fa07}
\bibinfo{author}{\bibfnamefont{A.}~\bibnamefont{Fasolino}},
  \bibinfo{author}{\bibfnamefont{J.~H.} \bibnamefont{Los}}, \bibnamefont{and}
  \bibinfo{author}{\bibfnamefont{M.~I.} \bibnamefont{Katsnelson}},
  \bibinfo{journal}{Nature Mater.} \textbf{\bibinfo{volume}{6}},
  \bibinfo{pages}{858} (\bibinfo{year}{2007}).

\bibitem[{\citenamefont{Meyer et~al.}(2007)\citenamefont{Meyer, Geim,
  Katsnelson, Novoselov, Booth, and Roth}}]{me07}
\bibinfo{author}{\bibfnamefont{J.~C.} \bibnamefont{Meyer}},
  \bibinfo{author}{\bibfnamefont{A.~K.} \bibnamefont{Geim}},
  \bibinfo{author}{\bibfnamefont{M.~I.} \bibnamefont{Katsnelson}},
  \bibinfo{author}{\bibfnamefont{K.~S.} \bibnamefont{Novoselov}},
  \bibinfo{author}{\bibfnamefont{T.~J.} \bibnamefont{Booth}}, \bibnamefont{and}
  \bibinfo{author}{\bibfnamefont{S.}~\bibnamefont{Roth}},
  \bibinfo{journal}{Nature} \textbf{\bibinfo{volume}{446}}, \bibinfo{pages}{60}
  (\bibinfo{year}{2007}).

\bibitem[{\citenamefont{de~Andres
  et~al.}(2012{\natexlab{a}})\citenamefont{de~Andres, Guinea, and
  Katsnelson}}]{an12}
\bibinfo{author}{\bibfnamefont{P.~L.} \bibnamefont{de~Andres}},
  \bibinfo{author}{\bibfnamefont{F.}~\bibnamefont{Guinea}}, \bibnamefont{and}
  \bibinfo{author}{\bibfnamefont{M.~I.} \bibnamefont{Katsnelson}},
  \bibinfo{journal}{Phys. Rev. B} \textbf{\bibinfo{volume}{86}},
  \bibinfo{pages}{144103} (\bibinfo{year}{2012}{\natexlab{a}}).

\bibitem[{\citenamefont{Evans and Rawicz}(1990)}]{ev90}
\bibinfo{author}{\bibfnamefont{E.}~\bibnamefont{Evans}} \bibnamefont{and}
  \bibinfo{author}{\bibfnamefont{W.}~\bibnamefont{Rawicz}},
  \bibinfo{journal}{Phys. Rev. Lett.} \textbf{\bibinfo{volume}{64}},
  \bibinfo{pages}{2094} (\bibinfo{year}{1990}).

\bibitem[{\citenamefont{Safran}(1994)}]{sa94}
\bibinfo{author}{\bibfnamefont{S.~A.} \bibnamefont{Safran}},
  \emph{\bibinfo{title}{Statistical Thermodynamics of Surfaces, Interfaces, and
  Membranes}} (\bibinfo{publisher}{Addison Wesley}, \bibinfo{address}{New
  York}, \bibinfo{year}{1994}).

\bibitem[{\citenamefont{Cerda and Mahadevan}(2003)}]{ce03}
\bibinfo{author}{\bibfnamefont{E.}~\bibnamefont{Cerda}} \bibnamefont{and}
  \bibinfo{author}{\bibfnamefont{L.}~\bibnamefont{Mahadevan}},
  \bibinfo{journal}{Phys. Rev. Lett.} \textbf{\bibinfo{volume}{90}},
  \bibinfo{pages}{074302} (\bibinfo{year}{2003}).

\bibitem[{\citenamefont{Wong and Pellegrino}(2006)}]{wo06}
\bibinfo{author}{\bibfnamefont{Y.~W.} \bibnamefont{Wong}} \bibnamefont{and}
  \bibinfo{author}{\bibfnamefont{S.}~\bibnamefont{Pellegrino}},
  \bibinfo{journal}{J. Mech. Materials Struct.} \textbf{\bibinfo{volume}{1}},
  \bibinfo{pages}{3} (\bibinfo{year}{2006}).

\bibitem[{\citenamefont{Kirilenko et~al.}(2011)\citenamefont{Kirilenko,
  Dideykin, and Van~Tendeloo}}]{ki11}
\bibinfo{author}{\bibfnamefont{D.~A.} \bibnamefont{Kirilenko}},
  \bibinfo{author}{\bibfnamefont{A.~T.} \bibnamefont{Dideykin}},
  \bibnamefont{and}
  \bibinfo{author}{\bibfnamefont{G.}~\bibnamefont{Van~Tendeloo}},
  \bibinfo{journal}{Phys. Rev. B} \textbf{\bibinfo{volume}{84}},
  \bibinfo{pages}{235417} (\bibinfo{year}{2011}).

\bibitem[{\citenamefont{Nicholl et~al.}(2015)\citenamefont{Nicholl, Conley,
  Lavrik, Vlassiouk, Puzyrev, Sreenivas, Pantelides, and Bolotin}}]{ni15}
\bibinfo{author}{\bibfnamefont{R.~J.~T.} \bibnamefont{Nicholl}},
  \bibinfo{author}{\bibfnamefont{H.~J.} \bibnamefont{Conley}},
  \bibinfo{author}{\bibfnamefont{N.~V.} \bibnamefont{Lavrik}},
  \bibinfo{author}{\bibfnamefont{I.}~\bibnamefont{Vlassiouk}},
  \bibinfo{author}{\bibfnamefont{Y.~S.} \bibnamefont{Puzyrev}},
  \bibinfo{author}{\bibfnamefont{V.~P.} \bibnamefont{Sreenivas}},
  \bibinfo{author}{\bibfnamefont{S.~T.} \bibnamefont{Pantelides}},
  \bibnamefont{and} \bibinfo{author}{\bibfnamefont{K.~I.}
  \bibnamefont{Bolotin}}, \bibinfo{journal}{Nature Commun.}
  \textbf{\bibinfo{volume}{6}}, \bibinfo{pages}{8789} (\bibinfo{year}{2015}).

\bibitem[{\citenamefont{Ruiz-Vargas et~al.}(2011)\citenamefont{Ruiz-Vargas,
  Zhuang, Huang, van~der Zande, Garg, McEuen, Muller, Hennig, and Park}}]{ru11}
\bibinfo{author}{\bibfnamefont{C.~S.} \bibnamefont{Ruiz-Vargas}},
  \bibinfo{author}{\bibfnamefont{H.~L.} \bibnamefont{Zhuang}},
  \bibinfo{author}{\bibfnamefont{P.~Y.} \bibnamefont{Huang}},
  \bibinfo{author}{\bibfnamefont{A.~M.} \bibnamefont{van~der Zande}},
  \bibinfo{author}{\bibfnamefont{S.}~\bibnamefont{Garg}},
  \bibinfo{author}{\bibfnamefont{P.~L.} \bibnamefont{McEuen}},
  \bibinfo{author}{\bibfnamefont{D.~A.} \bibnamefont{Muller}},
  \bibinfo{author}{\bibfnamefont{R.~G.} \bibnamefont{Hennig}},
  \bibnamefont{and} \bibinfo{author}{\bibfnamefont{J.}~\bibnamefont{Park}},
  \bibinfo{journal}{Nano Lett.} \textbf{\bibinfo{volume}{11}},
  \bibinfo{pages}{2259} (\bibinfo{year}{2011}).

\bibitem[{\citenamefont{Kosmrlj and Nelson}(2013)}]{ko13}
\bibinfo{author}{\bibfnamefont{A.}~\bibnamefont{Kosmrlj}} \bibnamefont{and}
  \bibinfo{author}{\bibfnamefont{D.~R.} \bibnamefont{Nelson}},
  \bibinfo{journal}{Phys. Rev. E} \textbf{\bibinfo{volume}{88}},
  \bibinfo{pages}{012136} (\bibinfo{year}{2013}).

\bibitem[{\citenamefont{Kosmrlj and Nelson}(2014)}]{ko14}
\bibinfo{author}{\bibfnamefont{A.}~\bibnamefont{Kosmrlj}} \bibnamefont{and}
  \bibinfo{author}{\bibfnamefont{D.~R.} \bibnamefont{Nelson}},
  \bibinfo{journal}{Phys. Rev. E} \textbf{\bibinfo{volume}{89}},
  \bibinfo{pages}{022126} (\bibinfo{year}{2014}).

\bibitem[{\citenamefont{Ram\'irez and Herrero}(2017)}]{ra17}
\bibinfo{author}{\bibfnamefont{R.}~\bibnamefont{Ram\'irez}} \bibnamefont{and}
  \bibinfo{author}{\bibfnamefont{C.~P.} \bibnamefont{Herrero}},
  \bibinfo{journal}{Phys. Rev. B} \textbf{\bibinfo{volume}{95}},
  \bibinfo{pages}{045423} (\bibinfo{year}{2017}).

\bibitem[{\citenamefont{L\'opez-Pol\'in
  et~al.}(2017)\citenamefont{L\'opez-Pol\'in, Jaafar, Guinea, Rold\'an,
  G\'omez-Navarro, and G\'omez-Herrero}}]{lo17}
\bibinfo{author}{\bibfnamefont{G.}~\bibnamefont{L\'opez-Pol\'in}},
  \bibinfo{author}{\bibfnamefont{M.}~\bibnamefont{Jaafar}},
  \bibinfo{author}{\bibfnamefont{F.}~\bibnamefont{Guinea}},
  \bibinfo{author}{\bibfnamefont{R.}~\bibnamefont{Rold\'an}},
  \bibinfo{author}{\bibfnamefont{C.}~\bibnamefont{G\'omez-Navarro}},
  \bibnamefont{and}
  \bibinfo{author}{\bibfnamefont{J.}~\bibnamefont{G\'omez-Herrero}},
  \bibinfo{journal}{Carbon} \textbf{\bibinfo{volume}{124}}, \bibinfo{pages}{42}
  (\bibinfo{year}{2017}).

\bibitem[{\citenamefont{Pozzo et~al.}(2011)\citenamefont{Pozzo, Alf\`e,
  Lacovig, Hofmann, Lizzit, and Baraldi}}]{po11b}
\bibinfo{author}{\bibfnamefont{M.}~\bibnamefont{Pozzo}},
  \bibinfo{author}{\bibfnamefont{D.}~\bibnamefont{Alf\`e}},
  \bibinfo{author}{\bibfnamefont{P.}~\bibnamefont{Lacovig}},
  \bibinfo{author}{\bibfnamefont{P.}~\bibnamefont{Hofmann}},
  \bibinfo{author}{\bibfnamefont{S.}~\bibnamefont{Lizzit}}, \bibnamefont{and}
  \bibinfo{author}{\bibfnamefont{A.}~\bibnamefont{Baraldi}},
  \bibinfo{journal}{Phys. Rev. Lett.} \textbf{\bibinfo{volume}{106}},
  \bibinfo{pages}{135501} (\bibinfo{year}{2011}).

\bibitem[{\citenamefont{Woods et~al.}(2014)\citenamefont{Woods, Britnell,
  Eckmann, Ma, Lu, Guo, Lin, Yu, Cao, Gorbachev et~al.}}]{wo14}
\bibinfo{author}{\bibfnamefont{C.~R.} \bibnamefont{Woods}},
  \bibinfo{author}{\bibfnamefont{L.}~\bibnamefont{Britnell}},
  \bibinfo{author}{\bibfnamefont{A.}~\bibnamefont{Eckmann}},
  \bibinfo{author}{\bibfnamefont{R.~S.} \bibnamefont{Ma}},
  \bibinfo{author}{\bibfnamefont{J.~C.} \bibnamefont{Lu}},
  \bibinfo{author}{\bibfnamefont{H.~M.} \bibnamefont{Guo}},
  \bibinfo{author}{\bibfnamefont{X.}~\bibnamefont{Lin}},
  \bibinfo{author}{\bibfnamefont{G.~L.} \bibnamefont{Yu}},
  \bibinfo{author}{\bibfnamefont{Y.}~\bibnamefont{Cao}},
  \bibinfo{author}{\bibfnamefont{R.~V.} \bibnamefont{Gorbachev}},
  \bibnamefont{et~al.}, \bibinfo{journal}{Nature Phys.}
  \textbf{\bibinfo{volume}{10}}, \bibinfo{pages}{451} (\bibinfo{year}{2014}).

\bibitem[{\citenamefont{Amorim et~al.}(2016)\citenamefont{Amorim, Cortijo,
  de~Juan, Grushine, Guinea, Gutierrez-Rubio, Ochoa, Parente, Roldan, San-Jose
  et~al.}}]{am16}
\bibinfo{author}{\bibfnamefont{B.}~\bibnamefont{Amorim}},
  \bibinfo{author}{\bibfnamefont{A.}~\bibnamefont{Cortijo}},
  \bibinfo{author}{\bibfnamefont{F.}~\bibnamefont{de~Juan}},
  \bibinfo{author}{\bibfnamefont{A.~G.} \bibnamefont{Grushine}},
  \bibinfo{author}{\bibfnamefont{F.}~\bibnamefont{Guinea}},
  \bibinfo{author}{\bibfnamefont{A.}~\bibnamefont{Gutierrez-Rubio}},
  \bibinfo{author}{\bibfnamefont{H.}~\bibnamefont{Ochoa}},
  \bibinfo{author}{\bibfnamefont{V.}~\bibnamefont{Parente}},
  \bibinfo{author}{\bibfnamefont{R.}~\bibnamefont{Roldan}},
  \bibinfo{author}{\bibfnamefont{P.}~\bibnamefont{San-Jose}},
  \bibnamefont{et~al.}, \bibinfo{journal}{Phys. Rep.}
  \textbf{\bibinfo{volume}{617}}, \bibinfo{pages}{1} (\bibinfo{year}{2016}).

\bibitem[{\citenamefont{Shimojo et~al.}(2008)\citenamefont{Shimojo, Kalia,
  Nakano, and Vashishta}}]{sh08}
\bibinfo{author}{\bibfnamefont{F.}~\bibnamefont{Shimojo}},
  \bibinfo{author}{\bibfnamefont{R.~K.} \bibnamefont{Kalia}},
  \bibinfo{author}{\bibfnamefont{A.}~\bibnamefont{Nakano}}, \bibnamefont{and}
  \bibinfo{author}{\bibfnamefont{P.}~\bibnamefont{Vashishta}},
  \bibinfo{journal}{Phys. Rev. B} \textbf{\bibinfo{volume}{77}},
  \bibinfo{pages}{085103} (\bibinfo{year}{2008}).

\bibitem[{\citenamefont{de~Andres
  et~al.}(2012{\natexlab{b}})\citenamefont{de~Andres, Guinea, and
  Katsnelson}}]{an12b}
\bibinfo{author}{\bibfnamefont{P.~L.} \bibnamefont{de~Andres}},
  \bibinfo{author}{\bibfnamefont{F.}~\bibnamefont{Guinea}}, \bibnamefont{and}
  \bibinfo{author}{\bibfnamefont{M.~I.} \bibnamefont{Katsnelson}},
  \bibinfo{journal}{Phys. Rev. B} \textbf{\bibinfo{volume}{86}},
  \bibinfo{pages}{245409} (\bibinfo{year}{2012}{\natexlab{b}}).

\bibitem[{\citenamefont{Chechin et~al.}(2014)\citenamefont{Chechin, Dmitriev,
  Lobzenko, and Ryabov}}]{ch14}
\bibinfo{author}{\bibfnamefont{G.~M.} \bibnamefont{Chechin}},
  \bibinfo{author}{\bibfnamefont{S.~V.} \bibnamefont{Dmitriev}},
  \bibinfo{author}{\bibfnamefont{I.~P.} \bibnamefont{Lobzenko}},
  \bibnamefont{and} \bibinfo{author}{\bibfnamefont{D.~S.}
  \bibnamefont{Ryabov}}, \bibinfo{journal}{Phys. Rev. B}
  \textbf{\bibinfo{volume}{90}}, \bibinfo{pages}{045432}
  (\bibinfo{year}{2014}).

\bibitem[{\citenamefont{Herrero and Ram\'irez}(2009)}]{he09a}
\bibinfo{author}{\bibfnamefont{C.~P.} \bibnamefont{Herrero}} \bibnamefont{and}
  \bibinfo{author}{\bibfnamefont{R.}~\bibnamefont{Ram\'irez}},
  \bibinfo{journal}{Phys. Rev. B} \textbf{\bibinfo{volume}{79}},
  \bibinfo{pages}{115429} (\bibinfo{year}{2009}).

\bibitem[{\citenamefont{Cadelano et~al.}(2009)\citenamefont{Cadelano, Palla,
  Giordano, and Colombo}}]{ca09c}
\bibinfo{author}{\bibfnamefont{E.}~\bibnamefont{Cadelano}},
  \bibinfo{author}{\bibfnamefont{P.~L.} \bibnamefont{Palla}},
  \bibinfo{author}{\bibfnamefont{S.}~\bibnamefont{Giordano}}, \bibnamefont{and}
  \bibinfo{author}{\bibfnamefont{L.}~\bibnamefont{Colombo}},
  \bibinfo{journal}{Phys. Rev. Lett.} \textbf{\bibinfo{volume}{102}},
  \bibinfo{pages}{235502} (\bibinfo{year}{2009}).

\bibitem[{\citenamefont{Akatyeva and Dumitrica}(2012)}]{ak12}
\bibinfo{author}{\bibfnamefont{E.}~\bibnamefont{Akatyeva}} \bibnamefont{and}
  \bibinfo{author}{\bibfnamefont{T.}~\bibnamefont{Dumitrica}},
  \bibinfo{journal}{J. Chem. Phys.} \textbf{\bibinfo{volume}{137}},
  \bibinfo{pages}{234702} (\bibinfo{year}{2012}).

\bibitem[{\citenamefont{Lee et~al.}(2013)\citenamefont{Lee, Yoon, Hwang, Wang,
  and Ho}}]{le13}
\bibinfo{author}{\bibfnamefont{G.-D.} \bibnamefont{Lee}},
  \bibinfo{author}{\bibfnamefont{E.}~\bibnamefont{Yoon}},
  \bibinfo{author}{\bibfnamefont{N.-M.} \bibnamefont{Hwang}},
  \bibinfo{author}{\bibfnamefont{C.-Z.} \bibnamefont{Wang}}, \bibnamefont{and}
  \bibinfo{author}{\bibfnamefont{K.-M.} \bibnamefont{Ho}},
  \bibinfo{journal}{Appl. Phys. Lett.} \textbf{\bibinfo{volume}{102}},
  \bibinfo{pages}{021603} (\bibinfo{year}{2013}).

\bibitem[{\citenamefont{Shen et~al.}(2013)\citenamefont{Shen, Xu, and
  Zhang}}]{sh13}
\bibinfo{author}{\bibfnamefont{H.-S.} \bibnamefont{Shen}},
  \bibinfo{author}{\bibfnamefont{Y.-M.} \bibnamefont{Xu}}, \bibnamefont{and}
  \bibinfo{author}{\bibfnamefont{C.-L.} \bibnamefont{Zhang}},
  \bibinfo{journal}{Appl. Phys. Lett.} \textbf{\bibinfo{volume}{102}},
  \bibinfo{pages}{131905} (\bibinfo{year}{2013}).

\bibitem[{\citenamefont{Los et~al.}(2009)\citenamefont{Los, Katsnelson, Yazyev,
  Zakharchenko, and Fasolino}}]{lo09}
\bibinfo{author}{\bibfnamefont{J.~H.} \bibnamefont{Los}},
  \bibinfo{author}{\bibfnamefont{M.~I.} \bibnamefont{Katsnelson}},
  \bibinfo{author}{\bibfnamefont{O.~V.} \bibnamefont{Yazyev}},
  \bibinfo{author}{\bibfnamefont{K.~V.} \bibnamefont{Zakharchenko}},
  \bibnamefont{and} \bibinfo{author}{\bibfnamefont{A.}~\bibnamefont{Fasolino}},
  \bibinfo{journal}{Phys. Rev. B} \textbf{\bibinfo{volume}{80}},
  \bibinfo{pages}{121405} (\bibinfo{year}{2009}).

\bibitem[{\citenamefont{Ram\'irez et~al.}(2016)\citenamefont{Ram\'irez,
  Chac\'on, and Herrero}}]{ra16}
\bibinfo{author}{\bibfnamefont{R.}~\bibnamefont{Ram\'irez}},
  \bibinfo{author}{\bibfnamefont{E.}~\bibnamefont{Chac\'on}}, \bibnamefont{and}
  \bibinfo{author}{\bibfnamefont{C.~P.} \bibnamefont{Herrero}},
  \bibinfo{journal}{Phys. Rev. B} \textbf{\bibinfo{volume}{93}},
  \bibinfo{pages}{235419} (\bibinfo{year}{2016}).

\bibitem[{\citenamefont{Magnin et~al.}(2014)\citenamefont{Magnin, Foerster,
  Rabilloud, Calvo, Zappelli, and Bichara}}]{ma14}
\bibinfo{author}{\bibfnamefont{Y.}~\bibnamefont{Magnin}},
  \bibinfo{author}{\bibfnamefont{G.~D.} \bibnamefont{Foerster}},
  \bibinfo{author}{\bibfnamefont{F.}~\bibnamefont{Rabilloud}},
  \bibinfo{author}{\bibfnamefont{F.}~\bibnamefont{Calvo}},
  \bibinfo{author}{\bibfnamefont{A.}~\bibnamefont{Zappelli}}, \bibnamefont{and}
  \bibinfo{author}{\bibfnamefont{C.}~\bibnamefont{Bichara}},
  \bibinfo{journal}{J. Phys.: Condens. Matter} \textbf{\bibinfo{volume}{26}},
  \bibinfo{pages}{185401} (\bibinfo{year}{2014}).

\bibitem[{\citenamefont{Los et~al.}(2016)\citenamefont{Los, Fasolino, and
  Katsnelson}}]{lo16}
\bibinfo{author}{\bibfnamefont{J.~H.} \bibnamefont{Los}},
  \bibinfo{author}{\bibfnamefont{A.}~\bibnamefont{Fasolino}}, \bibnamefont{and}
  \bibinfo{author}{\bibfnamefont{M.~I.} \bibnamefont{Katsnelson}},
  \bibinfo{journal}{Phys. Rev. Lett.} \textbf{\bibinfo{volume}{116}},
  \bibinfo{pages}{015901} (\bibinfo{year}{2016}).

\bibitem[{\citenamefont{Gillan}(1988)}]{gi88}
\bibinfo{author}{\bibfnamefont{M.~J.} \bibnamefont{Gillan}},
  \bibinfo{journal}{Phil. Mag. A} \textbf{\bibinfo{volume}{58}},
  \bibinfo{pages}{257} (\bibinfo{year}{1988}).

\bibitem[{\citenamefont{Ceperley}(1995)}]{ce95}
\bibinfo{author}{\bibfnamefont{D.~M.} \bibnamefont{Ceperley}},
  \bibinfo{journal}{Rev. Mod. Phys.} \textbf{\bibinfo{volume}{67}},
  \bibinfo{pages}{279} (\bibinfo{year}{1995}).

\bibitem[{\citenamefont{Brito et~al.}(2015)\citenamefont{Brito, C\^andido, Hai,
  and Peeters}}]{br15}
\bibinfo{author}{\bibfnamefont{B.~G.~A.} \bibnamefont{Brito}},
  \bibinfo{author}{\bibfnamefont{L.}~\bibnamefont{C\^andido}},
  \bibinfo{author}{\bibfnamefont{G.-Q.} \bibnamefont{Hai}}, \bibnamefont{and}
  \bibinfo{author}{\bibfnamefont{F.~M.} \bibnamefont{Peeters}},
  \bibinfo{journal}{Phys. Rev. B} \textbf{\bibinfo{volume}{92}},
  \bibinfo{pages}{195416} (\bibinfo{year}{2015}).

\bibitem[{\citenamefont{Mounet and Marzari}(2005)}]{mo05}
\bibinfo{author}{\bibfnamefont{N.}~\bibnamefont{Mounet}} \bibnamefont{and}
  \bibinfo{author}{\bibfnamefont{N.}~\bibnamefont{Marzari}},
  \bibinfo{journal}{Phys. Rev. B} \textbf{\bibinfo{volume}{71}},
  \bibinfo{pages}{205214} (\bibinfo{year}{2005}).

\bibitem[{\citenamefont{Shao et~al.}(2012)\citenamefont{Shao, Wen, Melnik, Yao,
  Kawazoe, and Tian}}]{sh12}
\bibinfo{author}{\bibfnamefont{T.}~\bibnamefont{Shao}},
  \bibinfo{author}{\bibfnamefont{B.}~\bibnamefont{Wen}},
  \bibinfo{author}{\bibfnamefont{R.}~\bibnamefont{Melnik}},
  \bibinfo{author}{\bibfnamefont{S.}~\bibnamefont{Yao}},
  \bibinfo{author}{\bibfnamefont{Y.}~\bibnamefont{Kawazoe}}, \bibnamefont{and}
  \bibinfo{author}{\bibfnamefont{Y.}~\bibnamefont{Tian}}, \bibinfo{journal}{J.
  Chem. Phys.} \textbf{\bibinfo{volume}{137}}, \bibinfo{pages}{194901}
  (\bibinfo{year}{2012}).

\bibitem[{\citenamefont{Gao and Huang}(2014)}]{ga14}
\bibinfo{author}{\bibfnamefont{W.}~\bibnamefont{Gao}} \bibnamefont{and}
  \bibinfo{author}{\bibfnamefont{R.}~\bibnamefont{Huang}}, \bibinfo{journal}{J.
  Mech. Phys. Solids} \textbf{\bibinfo{volume}{66}}, \bibinfo{pages}{42}
  (\bibinfo{year}{2014}).

\bibitem[{\citenamefont{Feynman}(1972)}]{fe72}
\bibinfo{author}{\bibfnamefont{R.~P.} \bibnamefont{Feynman}},
  \emph{\bibinfo{title}{Statistical Mechanics}}
  (\bibinfo{publisher}{Addison-Wesley}, \bibinfo{address}{New York},
  \bibinfo{year}{1972}).

\bibitem[{\citenamefont{Kleinert}(1990)}]{kl90}
\bibinfo{author}{\bibfnamefont{H.}~\bibnamefont{Kleinert}},
  \emph{\bibinfo{title}{Path Integrals in Quantum Mechanics, Statistics and
  Polymer Physics}} (\bibinfo{publisher}{World Scientific},
  \bibinfo{address}{Singapore}, \bibinfo{year}{1990}).

\bibitem[{\citenamefont{Chandler and Wolynes}(1981)}]{ch81}
\bibinfo{author}{\bibfnamefont{D.}~\bibnamefont{Chandler}} \bibnamefont{and}
  \bibinfo{author}{\bibfnamefont{P.~G.} \bibnamefont{Wolynes}},
  \bibinfo{journal}{J. Chem. Phys.} \textbf{\bibinfo{volume}{74}},
  \bibinfo{pages}{4078} (\bibinfo{year}{1981}).

\bibitem[{\citenamefont{Herrero and Ram\'irez}(2014)}]{he14}
\bibinfo{author}{\bibfnamefont{C.~P.} \bibnamefont{Herrero}} \bibnamefont{and}
  \bibinfo{author}{\bibfnamefont{R.}~\bibnamefont{Ram\'irez}},
  \bibinfo{journal}{J. Phys.: Condens. Matter} \textbf{\bibinfo{volume}{26}},
  \bibinfo{pages}{233201} (\bibinfo{year}{2014}).

\bibitem[{\citenamefont{Los et~al.}(2005)\citenamefont{Los, Ghiringhelli,
  Meijer, and Fasolino}}]{lo05}
\bibinfo{author}{\bibfnamefont{J.~H.} \bibnamefont{Los}},
  \bibinfo{author}{\bibfnamefont{L.~M.} \bibnamefont{Ghiringhelli}},
  \bibinfo{author}{\bibfnamefont{E.~J.} \bibnamefont{Meijer}},
  \bibnamefont{and} \bibinfo{author}{\bibfnamefont{A.}~\bibnamefont{Fasolino}},
  \bibinfo{journal}{Phys. Rev. B} \textbf{\bibinfo{volume}{72}},
  \bibinfo{pages}{214102} (\bibinfo{year}{2005}).

\bibitem[{\citenamefont{Ghiringhelli
  et~al.}(2005{\natexlab{a}})\citenamefont{Ghiringhelli, Los, Fasolino, and
  Meijer}}]{gh05}
\bibinfo{author}{\bibfnamefont{L.~M.} \bibnamefont{Ghiringhelli}},
  \bibinfo{author}{\bibfnamefont{J.~H.} \bibnamefont{Los}},
  \bibinfo{author}{\bibfnamefont{A.}~\bibnamefont{Fasolino}}, \bibnamefont{and}
  \bibinfo{author}{\bibfnamefont{E.~J.} \bibnamefont{Meijer}},
  \bibinfo{journal}{Phys. Rev. B} \textbf{\bibinfo{volume}{72}},
  \bibinfo{pages}{214103} (\bibinfo{year}{2005}{\natexlab{a}}).

\bibitem[{\citenamefont{Zakharchenko et~al.}(2009)\citenamefont{Zakharchenko,
  Katsnelson, and Fasolino}}]{za09}
\bibinfo{author}{\bibfnamefont{K.~V.} \bibnamefont{Zakharchenko}},
  \bibinfo{author}{\bibfnamefont{M.~I.} \bibnamefont{Katsnelson}},
  \bibnamefont{and} \bibinfo{author}{\bibfnamefont{A.}~\bibnamefont{Fasolino}},
  \bibinfo{journal}{Phys. Rev. Lett.} \textbf{\bibinfo{volume}{102}},
  \bibinfo{pages}{046808} (\bibinfo{year}{2009}).

\bibitem[{\citenamefont{Ghiringhelli
  et~al.}(2005{\natexlab{b}})\citenamefont{Ghiringhelli, Los, Meijer, Fasolino,
  and Frenkel}}]{gh05b}
\bibinfo{author}{\bibfnamefont{L.~M.} \bibnamefont{Ghiringhelli}},
  \bibinfo{author}{\bibfnamefont{J.~H.} \bibnamefont{Los}},
  \bibinfo{author}{\bibfnamefont{E.~J.} \bibnamefont{Meijer}},
  \bibinfo{author}{\bibfnamefont{A.}~\bibnamefont{Fasolino}}, \bibnamefont{and}
  \bibinfo{author}{\bibfnamefont{D.}~\bibnamefont{Frenkel}},
  \bibinfo{journal}{Phys. Rev. Lett.} \textbf{\bibinfo{volume}{94}},
  \bibinfo{pages}{145701} (\bibinfo{year}{2005}{\natexlab{b}}).

\bibitem[{\citenamefont{Politano et~al.}(2012)\citenamefont{Politano, Marino,
  Campi, Far\'ias, Miranda, and Chiarello}}]{po12}
\bibinfo{author}{\bibfnamefont{A.}~\bibnamefont{Politano}},
  \bibinfo{author}{\bibfnamefont{A.~R.} \bibnamefont{Marino}},
  \bibinfo{author}{\bibfnamefont{D.}~\bibnamefont{Campi}},
  \bibinfo{author}{\bibfnamefont{D.}~\bibnamefont{Far\'ias}},
  \bibinfo{author}{\bibfnamefont{R.}~\bibnamefont{Miranda}}, \bibnamefont{and}
  \bibinfo{author}{\bibfnamefont{G.}~\bibnamefont{Chiarello}},
  \bibinfo{journal}{Carbon} \textbf{\bibinfo{volume}{50}}, \bibinfo{pages}{4903
  } (\bibinfo{year}{2012}).

\bibitem[{\citenamefont{Lambin}(2014)}]{la14}
\bibinfo{author}{\bibfnamefont{P.}~\bibnamefont{Lambin}},
  \bibinfo{journal}{Appl. Sci.} \textbf{\bibinfo{volume}{4}},
  \bibinfo{pages}{282} (\bibinfo{year}{2014}).

\bibitem[{\citenamefont{Sfyris et~al.}(2015)\citenamefont{Sfyris, Koukaras,
  Pugno, and Galiotis}}]{sf15}
\bibinfo{author}{\bibfnamefont{D.}~\bibnamefont{Sfyris}},
  \bibinfo{author}{\bibfnamefont{E.~N.} \bibnamefont{Koukaras}},
  \bibinfo{author}{\bibfnamefont{N.}~\bibnamefont{Pugno}}, \bibnamefont{and}
  \bibinfo{author}{\bibfnamefont{C.}~\bibnamefont{Galiotis}},
  \bibinfo{journal}{J. Appl. Phys.} \textbf{\bibinfo{volume}{118}},
  \bibinfo{pages}{075301} (\bibinfo{year}{2015}).

\bibitem[{\citenamefont{Memarian et~al.}(2015)\citenamefont{Memarian,
  Fereidoon, and Ganji}}]{me15}
\bibinfo{author}{\bibfnamefont{F.}~\bibnamefont{Memarian}},
  \bibinfo{author}{\bibfnamefont{A.}~\bibnamefont{Fereidoon}},
  \bibnamefont{and} \bibinfo{author}{\bibfnamefont{M.~D.} \bibnamefont{Ganji}},
  \bibinfo{journal}{Superlattices Microstruct.} \textbf{\bibinfo{volume}{85}},
  \bibinfo{pages}{348} (\bibinfo{year}{2015}).

\bibitem[{\citenamefont{Zou et~al.}(2016)\citenamefont{Zou, Ye, and
  Cao}}]{zo16}
\bibinfo{author}{\bibfnamefont{J.-H.} \bibnamefont{Zou}},
  \bibinfo{author}{\bibfnamefont{Z.-Q.} \bibnamefont{Ye}}, \bibnamefont{and}
  \bibinfo{author}{\bibfnamefont{B.-Y.} \bibnamefont{Cao}},
  \bibinfo{journal}{J. Chem. Phys.} \textbf{\bibinfo{volume}{145}},
  \bibinfo{pages}{134705} (\bibinfo{year}{2016}).

\bibitem[{\citenamefont{Anastasi et~al.}(2016)\citenamefont{Anastasi, Ritos,
  Cassar, and Borg}}]{an16}
\bibinfo{author}{\bibfnamefont{A.~A.} \bibnamefont{Anastasi}},
  \bibinfo{author}{\bibfnamefont{K.}~\bibnamefont{Ritos}},
  \bibinfo{author}{\bibfnamefont{G.}~\bibnamefont{Cassar}}, \bibnamefont{and}
  \bibinfo{author}{\bibfnamefont{M.~K.} \bibnamefont{Borg}},
  \bibinfo{journal}{Mol. Simul.} \textbf{\bibinfo{volume}{42}},
  \bibinfo{pages}{1502} (\bibinfo{year}{2016}).

\bibitem[{\citenamefont{Ghasemi and Rajabpour}(2017)}]{gh17}
\bibinfo{author}{\bibfnamefont{H.}~\bibnamefont{Ghasemi}} \bibnamefont{and}
  \bibinfo{author}{\bibfnamefont{A.}~\bibnamefont{Rajabpour}},
  \bibinfo{journal}{J. Phys. Conf. Series} \textbf{\bibinfo{volume}{785}},
  \bibinfo{pages}{012006} (\bibinfo{year}{2017}).

\bibitem[{\citenamefont{Tuckerman et~al.}(1992)\citenamefont{Tuckerman, Berne,
  and Martyna}}]{tu92}
\bibinfo{author}{\bibfnamefont{M.~E.} \bibnamefont{Tuckerman}},
  \bibinfo{author}{\bibfnamefont{B.~J.} \bibnamefont{Berne}}, \bibnamefont{and}
  \bibinfo{author}{\bibfnamefont{G.~J.} \bibnamefont{Martyna}},
  \bibinfo{journal}{J. Chem. Phys.} \textbf{\bibinfo{volume}{97}},
  \bibinfo{pages}{1990} (\bibinfo{year}{1992}).

\bibitem[{\citenamefont{Tuckerman and Hughes}(1998)}]{tu98}
\bibinfo{author}{\bibfnamefont{M.~E.} \bibnamefont{Tuckerman}}
  \bibnamefont{and} \bibinfo{author}{\bibfnamefont{A.}~\bibnamefont{Hughes}},
  in \emph{\bibinfo{booktitle}{Classical and Quantum Dynamics in Condensed
  Phase Simulations}}, edited by \bibinfo{editor}{\bibfnamefont{B.~J.}
  \bibnamefont{Berne}},
  \bibinfo{editor}{\bibfnamefont{G.}~\bibnamefont{Ciccotti}}, \bibnamefont{and}
  \bibinfo{editor}{\bibfnamefont{D.~F.} \bibnamefont{Coker}}
  (\bibinfo{publisher}{Word Scientific}, \bibinfo{address}{Singapore},
  \bibinfo{year}{1998}), p. \bibinfo{pages}{311}.

\bibitem[{\citenamefont{Martyna et~al.}(1999)\citenamefont{Martyna, Hughes, and
  Tuckerman}}]{ma99}
\bibinfo{author}{\bibfnamefont{G.~J.} \bibnamefont{Martyna}},
  \bibinfo{author}{\bibfnamefont{A.}~\bibnamefont{Hughes}}, \bibnamefont{and}
  \bibinfo{author}{\bibfnamefont{M.~E.} \bibnamefont{Tuckerman}},
  \bibinfo{journal}{J. Chem. Phys.} \textbf{\bibinfo{volume}{110}},
  \bibinfo{pages}{3275} (\bibinfo{year}{1999}).

\bibitem[{\citenamefont{Tuckerman}(2002)}]{tu02}
\bibinfo{author}{\bibfnamefont{M.~E.} \bibnamefont{Tuckerman}}, in
  \emph{\bibinfo{booktitle}{Quantum Simulations of Complex Many--Body Systems:
  From Theory to Algorithms}}, edited by
  \bibinfo{editor}{\bibfnamefont{J.}~\bibnamefont{Grotendorst}},
  \bibinfo{editor}{\bibfnamefont{D.}~\bibnamefont{Marx}}, \bibnamefont{and}
  \bibinfo{editor}{\bibfnamefont{A.}~\bibnamefont{Muramatsu}}
  (\bibinfo{publisher}{NIC}, \bibinfo{address}{FZ J\"ulich},
  \bibinfo{year}{2002}), p. \bibinfo{pages}{269}.

\bibitem[{\citenamefont{Herman et~al.}(1982)\citenamefont{Herman, Bruskin, and
  Berne}}]{he82}
\bibinfo{author}{\bibfnamefont{M.~F.} \bibnamefont{Herman}},
  \bibinfo{author}{\bibfnamefont{E.~J.} \bibnamefont{Bruskin}},
  \bibnamefont{and} \bibinfo{author}{\bibfnamefont{B.~J.} \bibnamefont{Berne}},
  \bibinfo{journal}{J. Chem. Phys.} \textbf{\bibinfo{volume}{76}},
  \bibinfo{pages}{5150} (\bibinfo{year}{1982}).

\bibitem[{\citenamefont{Herrero et~al.}(2006)\citenamefont{Herrero,
  Ram\'{\i}rez, and Hern\'andez}}]{he06}
\bibinfo{author}{\bibfnamefont{C.~P.} \bibnamefont{Herrero}},
  \bibinfo{author}{\bibfnamefont{R.}~\bibnamefont{Ram\'{\i}rez}},
  \bibnamefont{and} \bibinfo{author}{\bibfnamefont{E.~R.}
  \bibnamefont{Hern\'andez}}, \bibinfo{journal}{Phys. Rev. B}
  \textbf{\bibinfo{volume}{73}}, \bibinfo{pages}{245211}
  (\bibinfo{year}{2006}).

\bibitem[{\citenamefont{Herrero and Ram\'irez}(2011)}]{he11}
\bibinfo{author}{\bibfnamefont{C.~P.} \bibnamefont{Herrero}} \bibnamefont{and}
  \bibinfo{author}{\bibfnamefont{R.}~\bibnamefont{Ram\'irez}},
  \bibinfo{journal}{J. Chem. Phys.} \textbf{\bibinfo{volume}{134}},
  \bibinfo{pages}{094510} (\bibinfo{year}{2011}).

\bibitem[{\citenamefont{Ram\'irez et~al.}(2012)\citenamefont{Ram\'irez,
  Neuerburg, Fern\'andez-Serra, and Herrero}}]{ra12}
\bibinfo{author}{\bibfnamefont{R.}~\bibnamefont{Ram\'irez}},
  \bibinfo{author}{\bibfnamefont{N.}~\bibnamefont{Neuerburg}},
  \bibinfo{author}{\bibfnamefont{M.~V.} \bibnamefont{Fern\'andez-Serra}},
  \bibnamefont{and} \bibinfo{author}{\bibfnamefont{C.~P.}
  \bibnamefont{Herrero}}, \bibinfo{journal}{J. Chem. Phys.}
  \textbf{\bibinfo{volume}{137}}, \bibinfo{pages}{044502}
  (\bibinfo{year}{2012}).

\bibitem[{\citenamefont{Ram\'irez et~al.}(2008)\citenamefont{Ram\'irez,
  Herrero, Hern\'andez, and Cardona}}]{ra08}
\bibinfo{author}{\bibfnamefont{R.}~\bibnamefont{Ram\'irez}},
  \bibinfo{author}{\bibfnamefont{C.~P.} \bibnamefont{Herrero}},
  \bibinfo{author}{\bibfnamefont{E.~R.} \bibnamefont{Hern\'andez}},
  \bibnamefont{and} \bibinfo{author}{\bibfnamefont{M.}~\bibnamefont{Cardona}},
  \bibinfo{journal}{Phys. Rev. B} \textbf{\bibinfo{volume}{77}},
  \bibinfo{pages}{045210} (\bibinfo{year}{2008}).

\bibitem[{\citenamefont{Fournier and Barbetta}(2008)}]{fo08}
\bibinfo{author}{\bibfnamefont{J.-B.} \bibnamefont{Fournier}} \bibnamefont{and}
  \bibinfo{author}{\bibfnamefont{C.}~\bibnamefont{Barbetta}},
  \bibinfo{journal}{Phys. Rev. Lett.} \textbf{\bibinfo{volume}{100}},
  \bibinfo{pages}{078103} (\bibinfo{year}{2008}).

\bibitem[{\citenamefont{Imparato}(2006)}]{im06}
\bibinfo{author}{\bibfnamefont{A.}~\bibnamefont{Imparato}},
  \bibinfo{journal}{J. Chem. Phys.} \textbf{\bibinfo{volume}{124}},
  \bibinfo{pages}{154714} (\bibinfo{year}{2006}).

\bibitem[{\citenamefont{Waheed and Edholm}(2009)}]{wa09}
\bibinfo{author}{\bibfnamefont{Q.}~\bibnamefont{Waheed}} \bibnamefont{and}
  \bibinfo{author}{\bibfnamefont{O.}~\bibnamefont{Edholm}},
  \bibinfo{journal}{Biophys. J.} \textbf{\bibinfo{volume}{97}},
  \bibinfo{pages}{2754} (\bibinfo{year}{2009}).

\bibitem[{\citenamefont{Chac\'on et~al.}(2015)\citenamefont{Chac\'on, Tarazona,
  and Bresme}}]{ch15}
\bibinfo{author}{\bibfnamefont{E.}~\bibnamefont{Chac\'on}},
  \bibinfo{author}{\bibfnamefont{P.}~\bibnamefont{Tarazona}}, \bibnamefont{and}
  \bibinfo{author}{\bibfnamefont{F.}~\bibnamefont{Bresme}},
  \bibinfo{journal}{J. Chem. Phys.} \textbf{\bibinfo{volume}{143}},
  \bibinfo{pages}{034706} (\bibinfo{year}{2015}).

\bibitem[{\citenamefont{Nicholl et~al.}(2017)\citenamefont{Nicholl, Lavrik,
  Vlassiouk, Srijanto, and Bolotin}}]{ni17}
\bibinfo{author}{\bibfnamefont{R.~J.~T.} \bibnamefont{Nicholl}},
  \bibinfo{author}{\bibfnamefont{N.~V.} \bibnamefont{Lavrik}},
  \bibinfo{author}{\bibfnamefont{I.}~\bibnamefont{Vlassiouk}},
  \bibinfo{author}{\bibfnamefont{B.~R.} \bibnamefont{Srijanto}},
  \bibnamefont{and} \bibinfo{author}{\bibfnamefont{K.~I.}
  \bibnamefont{Bolotin}}, \bibinfo{journal}{Phys. Rev. Lett.}
  \textbf{\bibinfo{volume}{118}}, \bibinfo{pages}{266101}
  (\bibinfo{year}{2017}).

\bibitem[{\citenamefont{Hahn et~al.}(2016)\citenamefont{Hahn, Melis, and
  Colombo}}]{ha16}
\bibinfo{author}{\bibfnamefont{K.~R.} \bibnamefont{Hahn}},
  \bibinfo{author}{\bibfnamefont{C.}~\bibnamefont{Melis}}, \bibnamefont{and}
  \bibinfo{author}{\bibfnamefont{L.}~\bibnamefont{Colombo}},
  \bibinfo{journal}{J. Phys. Chem. C} \textbf{\bibinfo{volume}{120}},
  \bibinfo{pages}{3026} (\bibinfo{year}{2016}).

\bibitem[{\citenamefont{Michel et~al.}(2015)\citenamefont{Michel, Costamagna,
  and Peeters}}]{mi15b}
\bibinfo{author}{\bibfnamefont{K.~H.} \bibnamefont{Michel}},
  \bibinfo{author}{\bibfnamefont{S.}~\bibnamefont{Costamagna}},
  \bibnamefont{and} \bibinfo{author}{\bibfnamefont{F.~M.}
  \bibnamefont{Peeters}}, \bibinfo{journal}{Phys. Status Solidi B}
  \textbf{\bibinfo{volume}{252}}, \bibinfo{pages}{2433} (\bibinfo{year}{2015}).

\bibitem[{\citenamefont{Landau and Lifshitz}(1980)}]{la80}
\bibinfo{author}{\bibfnamefont{L.~D.} \bibnamefont{Landau}} \bibnamefont{and}
  \bibinfo{author}{\bibfnamefont{E.~M.} \bibnamefont{Lifshitz}},
  \emph{\bibinfo{title}{Statistical Physics}} (\bibinfo{publisher}{Pergamon},
  \bibinfo{address}{Oxford}, \bibinfo{year}{1980}), \bibinfo{edition}{3rd} ed.

\bibitem[{\citenamefont{Herrero}(2002)}]{he02}
\bibinfo{author}{\bibfnamefont{C.~P.} \bibnamefont{Herrero}},
  \bibinfo{journal}{Phys. Rev. B} \textbf{\bibinfo{volume}{65}},
  \bibinfo{pages}{014112} (\bibinfo{year}{2002}).

\bibitem[{\citenamefont{Herrero and Ram\'{\i}rez}(1995)}]{he95}
\bibinfo{author}{\bibfnamefont{C.~P.} \bibnamefont{Herrero}} \bibnamefont{and}
  \bibinfo{author}{\bibfnamefont{R.}~\bibnamefont{Ram\'{\i}rez}},
  \bibinfo{journal}{Phys. Rev. B} \textbf{\bibinfo{volume}{51}},
  \bibinfo{pages}{16761} (\bibinfo{year}{1995}).

\bibitem[{\citenamefont{Landau and Lifshitz}(1965)}]{la65}
\bibinfo{author}{\bibfnamefont{L.~D.} \bibnamefont{Landau}} \bibnamefont{and}
  \bibinfo{author}{\bibfnamefont{E.~M.} \bibnamefont{Lifshitz}},
  \emph{\bibinfo{title}{Quantum Mechanics}} (\bibinfo{publisher}{Pergamon},
  \bibinfo{address}{Oxford}, \bibinfo{year}{1965}), \bibinfo{edition}{2nd} ed.

\bibitem[{\citenamefont{Gillan}(1990)}]{gi90}
\bibinfo{author}{\bibfnamefont{M.~J.} \bibnamefont{Gillan}}, in
  \emph{\bibinfo{booktitle}{Computer Modelling of Fluids, Polymers and
  Solids}}, edited by \bibinfo{editor}{\bibfnamefont{C.~R.~A.}
  \bibnamefont{Catlow}}, \bibinfo{editor}{\bibfnamefont{S.~C.}
  \bibnamefont{Parker}}, \bibnamefont{and}
  \bibinfo{editor}{\bibfnamefont{M.~P.} \bibnamefont{Allen}}
  (\bibinfo{publisher}{Kluwer}, \bibinfo{address}{Dordrecht},
  \bibinfo{year}{1990}), p. \bibinfo{pages}{155}.

\bibitem[{\citenamefont{Karssemeijer and Fasolino}(2011)}]{ka11}
\bibinfo{author}{\bibfnamefont{L.~J.} \bibnamefont{Karssemeijer}}
  \bibnamefont{and} \bibinfo{author}{\bibfnamefont{A.}~\bibnamefont{Fasolino}},
  \bibinfo{journal}{Surf. Sci.} \textbf{\bibinfo{volume}{605}},
  \bibinfo{pages}{1611} (\bibinfo{year}{2011}).

\bibitem[{\citenamefont{Wirtz and Rubio}(2004)}]{wi04}
\bibinfo{author}{\bibfnamefont{L.}~\bibnamefont{Wirtz}} \bibnamefont{and}
  \bibinfo{author}{\bibfnamefont{A.}~\bibnamefont{Rubio}},
  \bibinfo{journal}{Solid State Commun.} \textbf{\bibinfo{volume}{131}},
  \bibinfo{pages}{141} (\bibinfo{year}{2004}).

\end{thebibliography}
\end{document}